 \documentclass[final,5p]{elsarticle}
 \usepackage[utf8]{inputenc}
 \usepackage[T1]{fontenc}

 \usepackage{graphics}

\usepackage{amssymb}
\usepackage{makeidx}
\usepackage{graphicx}
\usepackage{color}
\usepackage{multicol}
\usepackage[bottom]{footmisc}
\usepackage[english]{babel}
\usepackage{amsmath}
\usepackage{xcolor}
\definecolor{link_blue}{RGB}{52,46,157}
\usepackage[colorlinks,citecolor=link_blue,linkcolor=link_blue,urlcolor=link_blue]{hyperref}

\biboptions{sort&compress}

\DeclareMathOperator{\rot}{rot}
\newcommand{\ri}{\mathrm{i}}
\newcommand{\ee}{\mathrm{e}}

\journal{NIM B}

\begin{document}

\begin{frontmatter}



\title{Parametric X-ray radiation in the Smith-Purcell geometry for
  non-destructive beam diagnostics}


\author[label6]{O. D. Skoromnik}
\ead{ods@mpi-hd.mpg.de}
\address[label6]{Max Planck Institute for Nuclear Physics,
  Saupfercheckweg 1, 69117 Heidelberg, Germany}

\author[label1,label2,label3]{I. D. Feranchuk\corref{cor1}}
\ead{ilya.feranchuk@tdtu.edu.vn}
\cortext[cor1]{Corresponding author}
\address[label1]{Atomic Molecular and Optical Physics Research Group,
  Advanced Institute of Materials Science, Ton Duc Thang University, 19
  Nguyen Huu Tho Str., Tan Phong Ward, District 7, Ho Chi Minh City,
  Vietnam}
\address[label2]{Faculty of Applied Sciences, Ton Duc Thang
  University, 19 Nguyen Huu Tho Str., Tan Phong Ward, District 7, Ho
  Chi Minh City, Vietnam}
\address[label3]{Belarusian State University, 4 Nezavisimosty Ave.,
  220030, Minsk, Belarus}

\author[label5]{D. V. Lu}
\address[label5]{Faculty of Physics, The University of Danang -
  University of Science and Education. B3 Building, 459 Ton Duc Thang,
  Lien Chieu, Da Nang, Vietnam}

\begin{abstract}
  We investigate parametric X-ray radiation (PXR) under condition of
  the extremely asymmetric diffraction, when the ultra-relativistic
  electron bunch is moving in \textit{vacuum} parallel to the
  crystal-vacuum interface, close to the crystal surface. This type of
  geometry coincides with the well known mechanism of generation of
  radiation, when the self-field of the particle beam interacts with
  the reflecting metal grating, namely the Smith-Purcell effect. We
  demonstrate that in this geometry the main contribution is given via
  a tail region of the beam distribution, which penetrates the crystal and
  X-rays are radiated along the normal to the crystal surface. We
  determine the electron beam characteristics, when this phenomenon
  can be observed. It is essential that in this geometry the majority
  of electrons does not undergo multiple scattering and
  consequently the characteristics of the particle beam are not
  changed, thus allowing the usage of the emitted X-rays for the
  purpose of non-destructive beam diagnostics, which can complement
  the traditional knife-edge method.
\end{abstract}


\end{frontmatter}


\section{Introduction}
\label{sec:introduction}

Parametric X-ray radiation (PXR) is generated when a charged particle
moves uniformly in a periodic medium \cite{PXR_Book_Feranchuk,
  J.Phys.France1985.46.1981}. The typical property of this type of
radiation is that it is emitted under the large angle to the velocity
of the charged particle. In addition, it is characterized by high
brightness, narrow spectral interval and possibility to uniformly tune
the frequency of the radiated photons. Moreover, the intensity of
radiation is relatively weakly dependent on the particle
energy. Furthermore, the large angle of the emitted photons allows one
to employ non-conventional geometries, which can lead to the
improvements of the various characteristics of the emitted radiation
\cite{PhysRevAccelBeams.21.014701}.

Recently, it was demonstrated \cite{SKOROMNIK201786} that the
intensity of the radiation can be significantly increased if the
grazing geometry under condition of the extremely asymmetric
diffraction of the emitted photons (PXR-EAD) is employed. However, in
that case the electrons were moving \textit{inside} a crystal,
parallel to the crystal vacuum interface. Consequently, in that
situation the effective length of the electron trajectory, which
contributes to the formation of PXR is bounded from above by the
multiple scattering on atoms of the medium, which the bunch of
electrons exhibits moving inside the crystal.

For this reason, it is essential to investigate the geometry, in which
the whole crystal length contributes to the intensity, but the
limiting factor of multiple electron scattering is removed. The most
natural way is to consider that the electron beam is moving
\textit{outside} of a crystal in vacuum, at a small distance to it,
but still parallel to the crystal-vacuum interface. This geometry
corresponds to the Smith-Purcell effect \cite{PhysRev.92.1069,
  vandenBerg:73}, but in the X-ray frequency range due to the
parametric radiation mechanism (PXR-SPG). In this case the electronic
density of a crystal corresponds to the metal surface grating,
interaction with which generates the radiation in optical or microwave
ranges \cite{PhysRevSTAB.7.070701, BARYSHEVSKY200692} and for X-rays
\cite{PhysRevLett.69.2523, PhysRevSTAB.18.052801}. For this reason,
the radiation field is formed due to the diffraction of the electron
beam self-field on the periodic electronic density of the
crystal. Consequently, the determination of the characteristics of
PXR-SPG, the discussion of its possible observation and applications
for the beam diagnostics \cite{PhysRevAccelBeams.21.052801} are the
main goals of the present work.

\section{Qualitative consideration}
\label{sec:qual-cons}

\begin{figure}[t]
  \centering
  \includegraphics[width=0.48\textwidth]{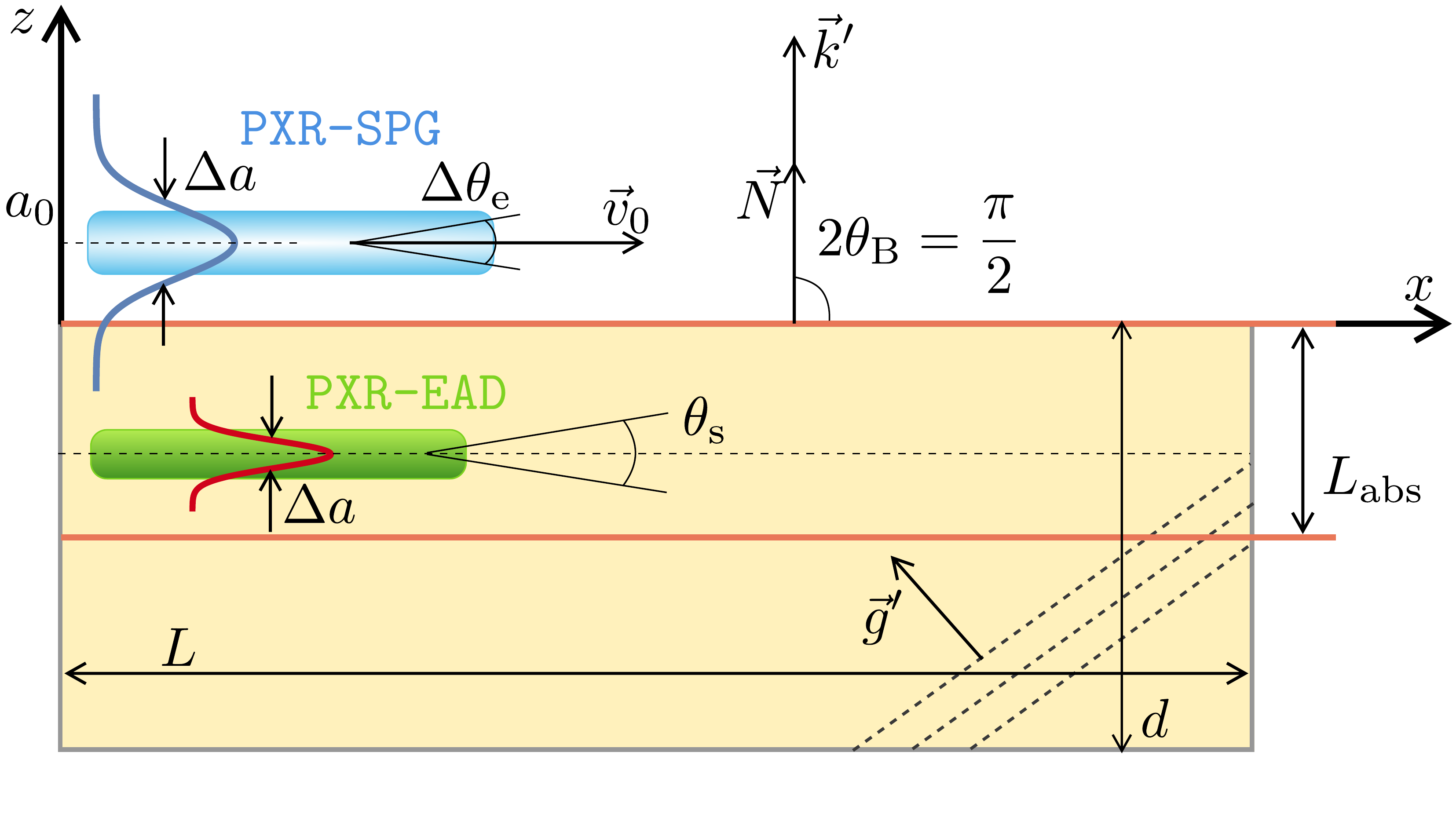}
  \caption{(Color online) The comparison of the grazing geometry of
    PXR-EAD, when an electron beam moves in a \textit{crystal} with
    the grazing geometry of PXR-SPG, when the main beam part
    propagates in \textit{vacuum} and only the halo of the beam
    travels inside a crystal. In both cases an electron beam
    propagates with the velocity $\vec v_{0}$ along the
    $\langle110\rangle$. The emitted radiation exits from a crystal in
    the direction $\vec k' = \omega \vec v_{0}/v_{0}^{2}+\vec g$ and
    is not absorbed. Here $\theta_{\mathrm{s}}$ is the mean square
    angle of multiple-electron scattering in a crystal. $a_{0}$ is the
    beam impact parameter, $\Delta a$ is the transverse spread of
    the beam due to its emittance and $\vec g'$ is the reciprocal lattice
    vector for planes where PXR is formed.}\label{fig:1}
\end{figure}

In order to discuss the qualitative peculiarities of the considered
phenomenon let us assume that the monocrystal plate of a thickness $d$
and a length $L$ is used to produce PXR-SPG. We also suppose that the
crystal length $L\gg d$, but the X-ray absorption length
$L_{\mathrm{abs}} < d$. In Fig.~\ref{fig:1} we demonstrate the
electron trajectories and compare the grazing geometry of PXR-EAD,
i.e., when an electron moves \textit{inside} the crystal parallel to
the crystal-vacuum interface with the grazing geometry of PXR-SPG when
an electron moves in \textit{vacuum} and also parallel to the
crystal-vacuum interface. In addition, we suppose that in the case of
PXR-SPG the electrons move closely to the crystal surface, such that
their own electromagnetic field penetrates the crystal. Following
Refs.~\cite{SKOROMNIK201786, J.Phys.France1985.46.1981} we now estimate
the number of radiated photons in both cases. Let the quantity
$Q_{\mathrm{PXR}}$ describes the number of photons emitted from the
unit length of an electron trajectory for the transition geometry
case. Then the total number of PXR photons is bounded by the X-ray
absorption length $L_{\mathrm{abs}}$ \cite{PXR_Book_Feranchuk} and equals to
\begin{align}
  N_{\mathrm{PXR}} = Q_{\mathrm{PXR}} L_{\mathrm{abs}}.\label{eq:1}
\end{align}
The actual expression for $Q_{\mathrm{PXR}}$ is not important and for
our qualitative estimation we mention only that $Q_{\mathrm{PXR}}$ is
independent of the crystal length under the condition
$L_{\mathrm{abs}} < d$ \cite{J.Phys.France1985.46.1981}.

In the case of PXR-EAD \cite{SKOROMNIK201786} the photons emitted from
the whole length $L$ of an electron trajectory contribute to the
formation of PXR, as they are not absorbed. Moreover, according to
Ref.~\cite{SKOROMNIK201786} the number of emitted photons
$N_{\mathrm{PXR-EAD}}$ in this situation significantly exceeds
$N_{\mathrm{PXR}}$. One can obtain the following estimation in this
case
\begin{align}
  N_{\mathrm{PXR-EAD}} = Q_{\mathrm{PXR-EAD}} L =
  10^{2}N_{\mathrm{PXR}}.\label{eq:2}
\end{align}

For the estimation of the number of photons $N_{\mathrm{PXR-SPG}}$ of
PXR-SPG we can use an analogous approach. First of all we notice that
in the Smith-Purcell geometry, due to the beam emittance, a part of an
electron beam (beam halo) is moving inside a crystal and generates
radiation according to PXR-EAD. At the same time the major part of the
beam is moving above the crystal surface and radiates according to the
Smith-Purcell effect ($N_{\mathrm{SP}}$). As a result the total number
of the emitted photons can be estimated in the following way
\begin{align}
  N_{\mathrm{PXR-SPG}} &= p_{\mathrm{i}} N_{\mathrm{PXR-EAD}} +
  p_{\mathrm{o}}N_{\mathrm{SP}},\label{eq:3}
  \\
  p_{\mathrm{i}} &+ p_{\mathrm{o}} = 1, \nonumber
\end{align}
where $p_{\mathrm{i}}$ is the probability that the electron of a beam
is moving inside the crystal and $p_{\mathrm{o}}$ is the corresponding
probability when an electron moves in vacuum. We mention here that
when the beam impact parameter $a_{0} \le 0$, $N_{\mathrm{PXR-SPG}}$
coincides with $N_{\mathrm{PXR-EAD}}$. Now let us estimate the
quantity $N_{\mathrm{SP}}$ in the ideal case of vanishing emittance
$\epsilon = 0$. In this situation, when crystal parameters are fixed
$N_{\mathrm{SP}}$ strongly depends on the impact parameter
$a_{0}$. Consequently,
\begin{align}
  N_{\mathrm{SP}} = Q_{\mathrm{SP}} L_{\mathrm{SP}}(a_{0}),\label{eq:4}
\end{align}
where we introduced the coherent length $L_{\mathrm{SP}}$ for this
process. This length is independent of the multiple electron
scattering (electrons are moving in vacuum). For its determination one
can use the following fact. It is well known
\cite{ginzburg2013theoretical} that the Fourier component of an
electromagnetic self-field of an electron, corresponding to the wave
length $\lambda$, in the direction perpendicular to the electron
velocity is located inside the region of an angular spread
$\theta_{\mathrm{tr}}$ and the characteristic size $a_{\mathrm{tr}}$,
for which one can write
\begin{align}
  \theta_{\mathrm{tr}} \approx \gamma^{-1}, \quad a_{\mathrm{tr}} \approx \frac{\lambda\gamma}{2\pi}, \quad \gamma = \frac{E}{mc^{2}}.\label{eq:5}
\end{align}

In the X-ray frequency range the characteristic size can reach the
value $a_{\mathrm{tr}} = 10^{-5}$ cm, when the particle energy
$E \approx 10^{3}$ MeV.

For the following we assume that the bunch of electrons is moving in
vacuum parallel to the crystal-vacuum interface under the distance
$a_{0}$, which is smaller than $a_{\mathrm{tr}}$, i.e.,
$a_{0}<a_{\mathrm{tr}}$. If this condition is not fulfilled the
intensity of the emitted radiation is exponentially suppressed, see
below. This bunch of electrons has a transverse size $\Delta a$,
$\Delta a < a_{\mathrm{tr}}$ and the angular spread
$\Delta \theta_{\mathrm{e}}$ that correspond to the natural emittance
$\epsilon = \Delta a \Delta \theta_{\mathrm{e}}$, see
Fig.~\ref{fig:1}. Consequently, the coherent length $L_{\mathrm{SP}}$
can be defined as the length of an electron trajectory when the
electron self-field still penetrates the crystal and therefore can
diffract on the electronic density of its atoms. In the ideal case of
vanishing transverse spread $\Delta a = 0$ this interaction takes
place along the whole crystal length that is
\begin{align}
  L_{\mathrm{SP}}(a_{0}) = L, \quad a_{0} < a_{\mathrm{tr}}, \quad
  \Delta a = 0. \label{eq:6}
\end{align}
\begin{figure*}[t]
  \centering
  \includegraphics[width=0.42\textwidth]{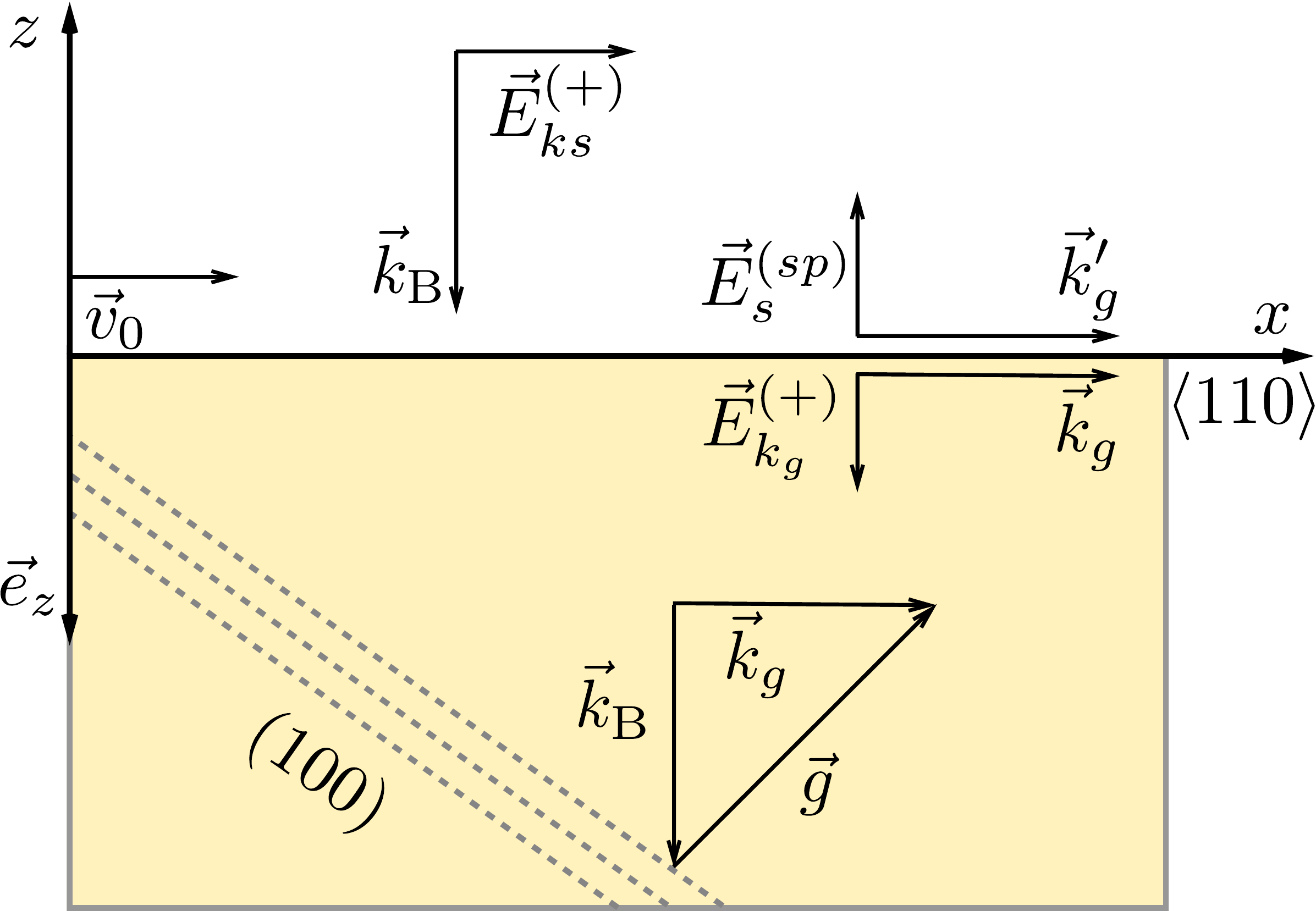}
  \includegraphics[width=0.42\textwidth]{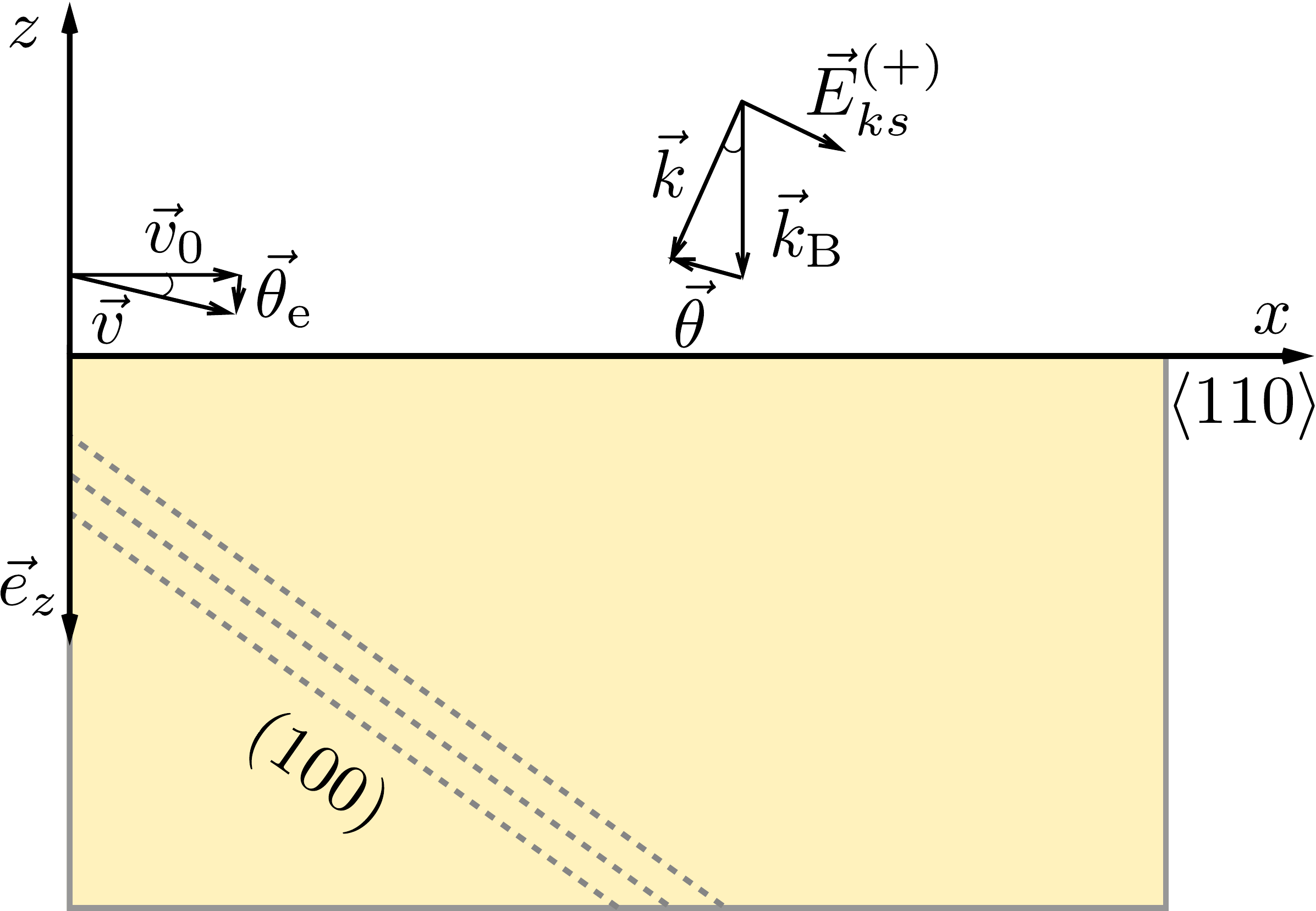}
  \caption{(Color online) Left pane: The grazing geometry of PXR-SPG in
    the ideal case, when the Wulff–Bragg's condition for the emitted
    photons is fulfilled and propagation directions of incident,
    diffracted and diffracted-reflected waves. The angle $\theta_{0}$
    is the angle between $\vec v_{0}$ and the $x$-axis and in the
    ideal case $\theta_{0} = 0$. Right pane: Non ideal case. The
    deviation from the Wulff–Bragg's condition and the variation of
    the velocity of the center of the beam are described by the
    vectors $\vec \theta$ and $\vec\theta_{\mathrm{e}}$
    respectively. In this pane the diffracted and diffracted-reflected
    waves are not shown. In both panes the incident wave
    $\vec E_{\vec k s}^{(+)}$ describes the emitted PXR field in
    accordance with the reciprocity theorem
    Eq.~(\ref{eq:28}).}\label{fig:2}
\end{figure*}

However, in the realistic situation of a non-vanishing emittance the
actual value of the coherent length is substantially restricted by the
actual angular and transverse spreads of experimentally available
electron bunches. As follows from Fig.~\ref{fig:1} one can write the
following estimation
\begin{align}
  L_{\mathrm{SP}}
  = \frac{a_{\mathrm{tr}}}{\Delta\theta_{\mathrm{e}}}
  \leq \frac{a_{\mathrm{tr}}^{2}}{\Delta a \Delta\theta_{\mathrm{e}}}
  = \frac{\lambda^{2}\gamma^{2}}{4\pi^{2}\epsilon},
  \quad a_{0} < \frac{\lambda\gamma}{2\pi}. \label{eq:7}
\end{align}

For the numerical considerations we employ the characteristics of the
Mainz microtron MAMI \cite{Brenzinger1997,Lauth2006}. As a result for
the radiation wavelength $\lambda = 3\times 10^{-8}$ cm, the emittance
$\epsilon = 3\times 10^{-7}$ cm and $\gamma = 2\times 10^{3}$ the
coherent length reaches
\begin{align}
  L_{\mathrm{SP}} \approx 10^{-3}\text{ cm},\label{eq:8}
\end{align}
which is comparable with $L_{\mathrm{abs}}$. These estimations are
justified for the photons that are emitted in the cone with the angle
$\theta < \theta_{\mathrm{tr}}$.

Let us now estimate the number of photons $Q_{\mathrm{SP}}$ emitted
from the unit length of a trajectory for the SP and compare its value
with the $Q_{\mathrm{PXR}}$. For this we notice
\cite{PXR_Book_Feranchuk, BARYSHEVSKY1986306,
  J.Phys.France1983.44.913} that the angular spread of the PXR is
defined via a parameter
\begin{align}
  \theta_{\mathrm{ph}} = \sqrt{\gamma^{-2} + \theta_{\mathrm{s}}^{2} + |\chi_{0}'|},\label{eq:9}
\end{align}
where $\theta_{\mathrm{s}}$ is the mean square of the electron
scattering angle and $\chi_{0}'$ is the real part of the dielectric
susceptibility of a crystal. Moreover, according to
Ref.~\cite{PXR_Book_Feranchuk} the $Q_{\mathrm{PXR}}$ is defined as
\begin{align}
  Q_{\mathrm{PXR}} \approx \left(\frac{\theta_{\mathrm{D}}}{\theta_{\mathrm{ph}}}\right)^{4} A. \label{eq:10}
\end{align}
Here $\theta_{\mathrm{D}}\approx \theta_{\mathrm{ph}}$ is the detector
aperture and $A$ is the quantity, which depends only on the parameters
of the crystal.

Analogously, for SP the detector aperture angle
$\theta_{\mathrm{D}}$ is limited by $\theta_{\mathrm{tr}}$
and for the angular width of the SP peak one can write
\begin{align}
  \theta_{\mathrm{ph}}^{(0)} = \sqrt{\gamma^{-2} + |\chi_{0}'|}, \label{eq:11}
\end{align}
where we omitted the electron scattering angle, since in the SP
case there is no multiple electron scattering as electrons are moving
in vacuum. Consequently, we can write an analogous expression to Eq.~(\ref{eq:11}), but for the SP
\begin{align}
  Q_{\mathrm{SP}} = \left(\frac{\theta_{\mathrm{tr}}}{\theta_{\mathrm{ph}}^{(0)}}\right)^{4} A = \left(\frac{1}{\gamma\theta_{\mathrm{ph}}^{(0)}}\right)^{4} A. \label{eq:12}
\end{align}
Consequently, the combination of Eqs.~(\ref{eq:3}), (\ref{eq:7}) and (\ref{eq:12}) yields
\begin{align}
  N_{\mathrm{SP}} = \frac{1}{\gamma^{2}(\theta_{\mathrm{ph}}^{(0)})^{4}}\frac{\lambda^{2}}{4\pi^{2}\epsilon L_{\mathrm{abs}}}N_{\mathrm{PXR}}. \label{eq:13}
\end{align}

The estimation according to Eq.~(\ref{eq:13}) for the MAMI microtron
demonstrates that the total number of emitted photons of SP is
significantly lower than the corresponding number for PXR and PXR-EAD:
\begin{align}
  N_{\mathrm{SP}} \approx 10^{-2}N_{\mathrm{PXR}} \approx
  10^{-4}N_{\mathrm{PXR-EAD}}.\label{eq:14}
\end{align}

Consequently, even if the small number of electrons from a tail region
of the beam distribution appears inside the crystal, exactly these
electrons will determine the total intensity of PXR-SPG
($p_{\mathrm{i}} \sim 10^{-2}p_{\mathrm{o}}$)
\begin{align}
  N_{\mathrm{PXR-SPG}} \approx p_{\mathrm{i}}N_{\mathrm{PXR-EAD}}.\label{eq:15}
\end{align}

Formulas (\ref{eq:11}) and (\ref{eq:13}) allow one to determine the
required normalized emittance for the number of emitted quanta
$N_{\mathrm{SP}}$ to be comparable with $p_{\mathrm{i}}N_{\mathrm{PXR-EAD}} =
N_{\mathrm{PXR}}$
\begin{align}
  \gamma\epsilon < \frac{1}{ \gamma(
  |\chi_{0}'|)^{2}}\frac{\lambda^{2}}{4\pi^{2} L_{\mathrm{abs}}}
  .\label{eq:16}
\end{align}

For the radiation wavelength $\lambda = 3\times 10^{-8}$ cm,
$\gamma = 2\times 10^{3}$ and Si crystal it leads to the condition
\begin{align}
  \gamma\epsilon \leq 10^{-8}\ \mathrm{cm} \times
  \mathrm{rad},\label{eq:17} 
\end{align}
which is unreachable for modern accelerators.

The above qualitative analysis demonstrates that the intensity of
PXR-SPG is mainly defined by the radiation of the electrons from the
beam halo, which move inside a crystal. As a result one can exploit
this feature for the non-destructive diagnostics of the electron beam.

In order to confirm these estimations we provide a rigorous
calculation of the spectral angular distribution of
$N_{\mathrm{PXR-SPG}}$ and the total number of emitted quanta based on
the dynamical diffraction theory, which we discuss in the subsequent
sections.

\section{Spectral--angular distribution and integral intensity of PXR-SPG}
\label{sec:spectr-angul-distr}

In this section we will apply the dynamical theory of diffraction in
order to calculate the intensity of PXR-SPG. However, we would like to
notice that we use a special approach to the solution of a radiation
problem, which was developed earlier in
Refs.~\cite{PXR_Book_Feranchuk, J.Phys.France1983.44.913,
  doi:10.1143/JPSJ.69.3462, baryshevsky2012high}. In the majority of
works, devoted to this problem when the interface between two media
exists, the Maxwell equations are solved for each media
independently. Then the resulting solution is represented as a linear
combination of two linearly independent solutions $\vec E^{(1)}_{i}$
and $\vec E^{(2)}_{i}$ of the homogeneous equations and the solution
$\vec E^{(\mathrm{in})}_{i}$ of the inhomogeneous ones. The solution
$\vec E^{(\mathrm{in})}_{i}$ is given via a current, formed by a
charged particle in the corresponding medium. Here the index $i$
numerates different media. Consequently, in this approach we can write
down the general solution for the electromagnetic field in each medium
as a linear combination
\begin{align}
  \vec E_{i} (\vec r,t) &= C^{(1)}_{i}  \vec E^{(1)}_{i} (\vec r,t) + C^{(2)}_{i}  \vec E^{(2)}_{i} (\vec r,t)\nonumber
  \\
  &+ \vec E^{(\mathrm{in})}_{i} (\vec r,t).\label{eq:18}
\end{align}

The electromagnetic field $\vec E_{i} (\vec r,t)$ is then plugged in
into the boundary conditions of electrodynamics, which leads to the
system of linear inhomogeneous equations for the coefficients
$C^{(1,2)}_{i}$. The solutions of these equations are expressed
through $\vec E^{(\mathrm{in})}_{i} (\vec r,t)$ and define the
intensity of the radiation at a large distance from the
crystal. However, the construction of a particular solution of
inhomogeneous equations is rather a difficult step.

In our work we will employ a different approach
\cite{PXR_Book_Feranchuk, J.Phys.France1983.44.913,
  doi:10.1143/JPSJ.69.3462, baryshevsky2012high}, which is based on
the solution of the homogeneous equations and a Green function in an
inhomogeneous medium in the whole space. Thus, for reader's
convenience we quickly revise below this procedure.

We are interested in the solution of the Maxwell's equations
\begin{align}
  \rot\rot \vec E(\vec r,t) + \frac{1}{c^{2}}\frac{\partial^{2}\vec
  D(\vec r,t)}{\partial t^{2}} = -\frac{4\pi}{c^{2}}\frac{\partial
  \vec j(\vec r,t)}{\partial t}, \label{eq:19}
\end{align}
or introducing the Fourier components
\begin{align}
  &\rot\rot\vec E (\vec r,\omega) - \frac{\omega^2}{c^2} \vec D (\vec
    r,\omega) = \mathrm{i} \omega \frac{4\pi}{c^2} \vec j (\vec
    r,\omega),\label{eq:20}
  \\
  &\vec j (\vec r,\omega) = \frac{e_{0}}{2 \pi}\int dt \vec v_{0}(t)
    \delta [ \vec r - \vec r_{0}(t)] \mathrm{e}^{\mathrm{i}\omega
    t}.\label{eq:21}
\end{align}

The tensor of dielectric permittivity
$\epsilon_{\alpha\beta}(\vec r,\vec r_{1},\omega)$ of a medium relates
the components of the induction vector $D_{\alpha}(\vec r,\omega)$
with the components $E_{\alpha}(\vec r,\omega)$ of the electromagnetic
field strength:
\begin{align}
  D_{\alpha}(\vec r, \omega) = \int d\vec r_{1}
  \epsilon_{\alpha\beta}(\vec r, \vec r_{1},\omega)E_{\beta}(\vec
  r_{1},\omega). \label{eq:22}
\end{align}
The dielectric permittivity tensor takes into account the boundaries
between media. In our case of vacuum and a crystal this means that in
vacuum
$\epsilon_{\alpha\beta}(\vec r, \vec r_{1},\omega) =
\delta_{\alpha\beta}\delta(\vec r - \vec r_{1})$ and
\begin{align*}
  \epsilon_{\alpha \beta}(t, \vec r, \vec r') = \sum_{\vec g}\int
d\omega\int d\vec k \epsilon_{\alpha \beta}(\vec k, \vec k + \vec g, \omega)
  \\
  \mspace{180mu}\times\mathrm{e}^{\ri\vec k\cdot(\vec r - \vec r') - \ri \omega
t - \ri\vec g\cdot\vec r}
\end{align*}
in the crystal respectively. The Fourier components
\begin{align*}
  \epsilon_{\alpha \beta}(\vec k, \vec k + \vec g, \omega)
\end{align*}
in this expression correspond to the periodic distribution of the
electronic density and are determined by the reciprocal lattice
vectors $\vec g$.

We now define the Green function of the Maxwell's equations
\begin{align}
  \varepsilon_{\alpha\beta\gamma}
  &\varepsilon_{\gamma\mu\nu} \frac{\partial^{2}}{\partial
    x_{\beta}\partial x_{\mu}}
    G_{\nu\lambda}(\vec r, \vec r', \omega) \label{eq:23}
  \\
  &-\frac{\omega^{2}}{c^{2}}\int d\vec
    r_{1}\epsilon_{\alpha\beta}(\vec r_{1}, \vec r',\omega)
    G_{\beta\lambda}(\vec r, \vec r', \omega) = \delta_{\alpha\lambda}
    \delta(\vec r - \vec r') \nonumber
\end{align}
where ${\sf\varepsilon_{\alpha \beta \gamma}}$ is a Levi-Civita
tensor. The Greek indices in Eq.~(\ref{eq:23}) are running as
$1\ldots3$ and a summation over repeated indices is understood. We
consider only the spontaneous emission, which vanishes in the absence
of the current. Consequently, with the help of the introduced Green
function one can write down the expression for the electromagnetic
field
\begin{align}
  E_{\alpha} (\vec r,\omega) = \ri \omega \frac{4
\pi}{c^2}\int d\vec r' G_{\alpha \beta}(\vec r,\vec r',\omega)j_{\beta} (\vec r',\omega).\label{eq:24}
\end{align}
\begin{figure}[t]
  \centering
  \includegraphics[width=0.48\textwidth]{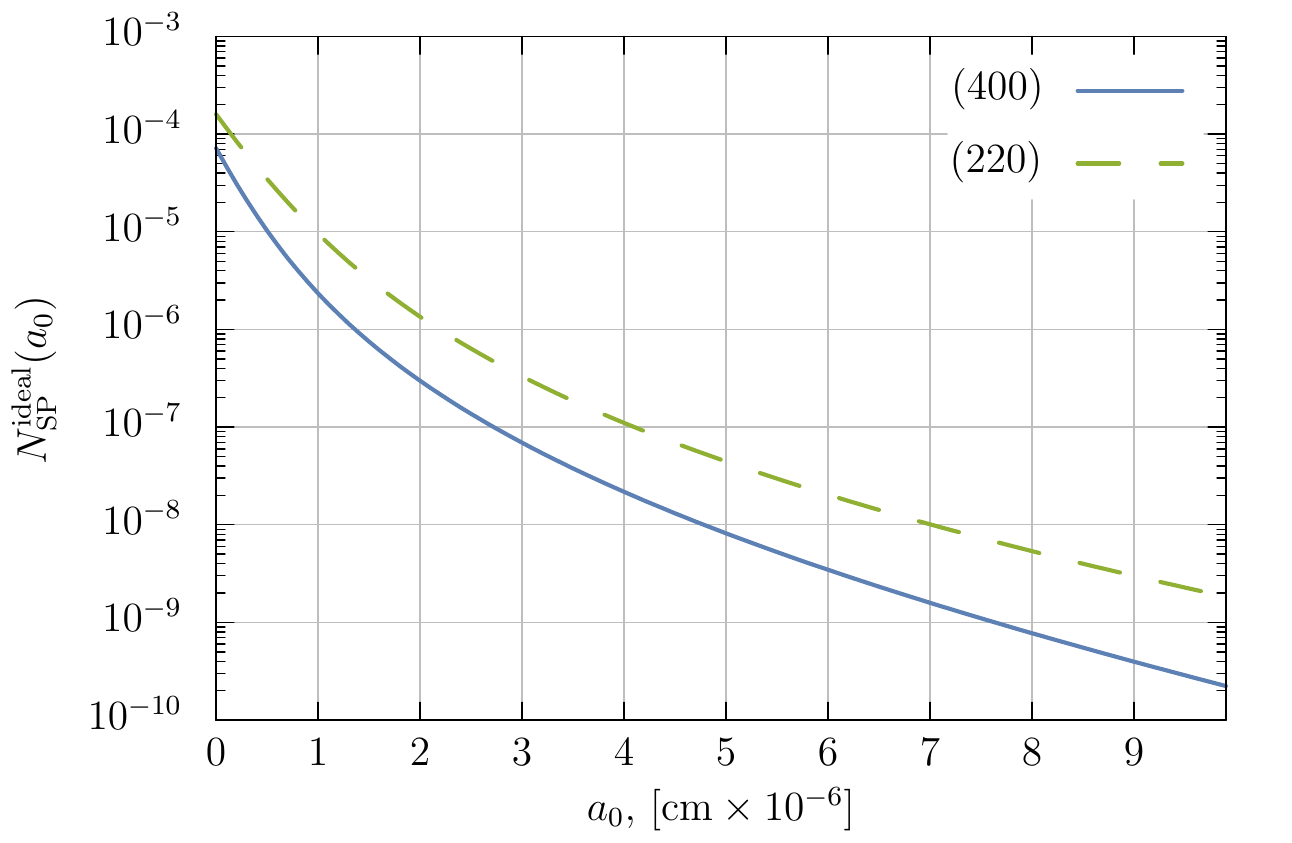}
  \caption{(Color online) The dependence of the Smith-Purcell photons
    $N_{\mathrm{SP}}^{\mathrm{ideal}}$ on the distance $a_{0}$ between
    the center of the electron beam and the crystal surface for the
    MAMI experimental facility in the ideal case of a vanishing
    emittance for the $(220)$ and $(400)$ reflections. The parameters
    of the reflections are given via Eqs.~(\ref{eq:61}) and
    (\ref{eq:62}). The crystal length $L = 1\,\mathrm{cm}$ and the
    electron beam energy $900$ MeV.}\label{fig:9}
\end{figure}
\begin{figure}[t]
  \centering
  \includegraphics[width=0.48\textwidth]{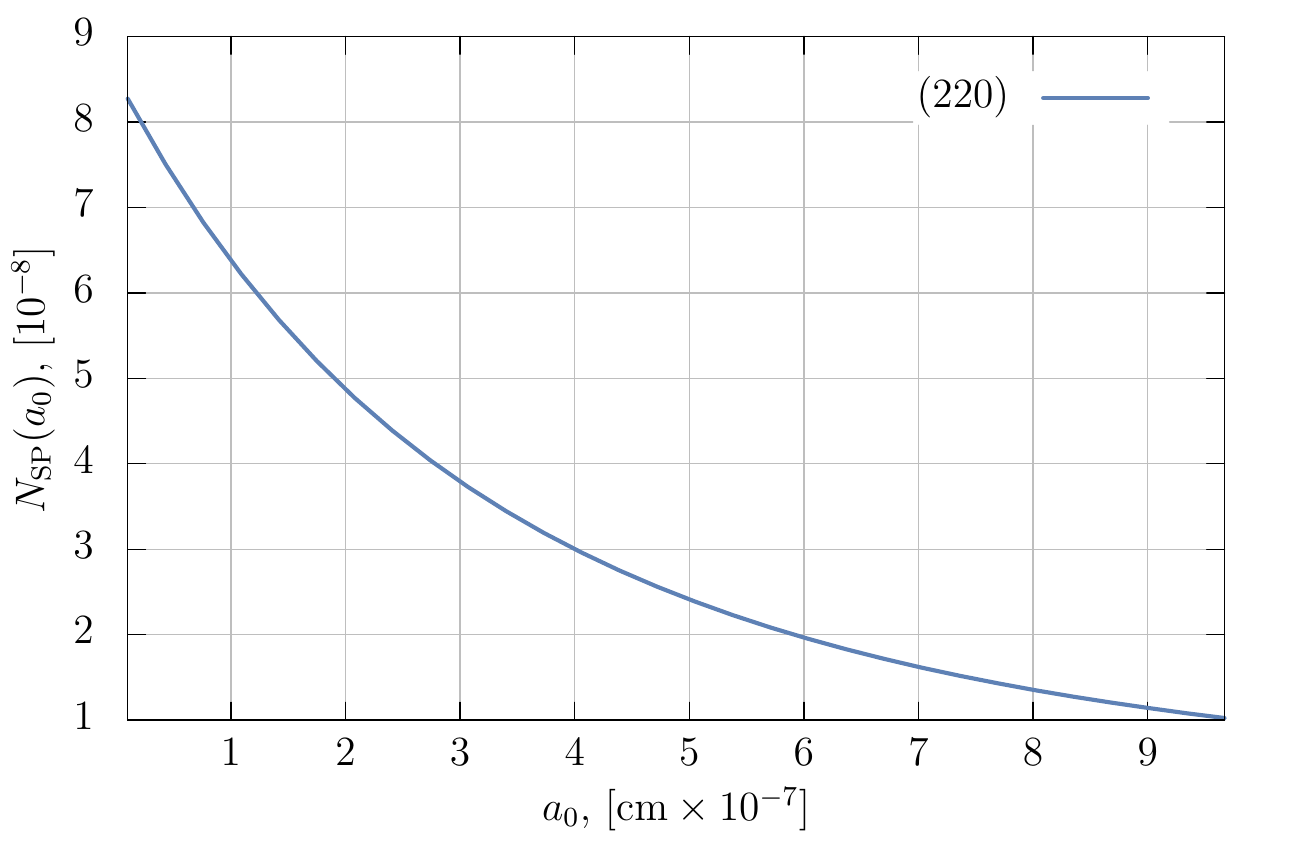}
  \caption{(Color online) The dependence of the Smith-Purcell photons
    $N_{\mathrm{SP}}$ on the distance $a_{0}$ between the center of
    the electron beam and the crystal surface for the MAMI
    experimental facility and the $(220)$ reflection. Here
    $\Delta a = a_{0} / 2$ and changes with $a_{0}$. The parameters of
    the reflection $(220)$ are given via Eq.~(\ref{eq:62}). The
    crystal length $L = 1\,\mathrm{cm}$, the angle
    $\theta_{0} = 10^{-5}\,\mathrm{rad}$ and the electron beam energy
    $900$ MeV.}\label{fig:4}
\end{figure}

In order to determine the radiation field in the far zone, where a
detector of X-ray photons is located we need to perform an asymptotic
expansion in the limit $r\gg r'$ in the Green function
\cite{PXR_Book_Feranchuk}. The resulting expression then reads
\begin{align}
  G_{\alpha \beta} (\vec r,\vec r',\omega)  &\approx
\frac{\mathrm{e}^{\ri kr}}{4 \pi r} \sum_{s=1,2}
  e_{s\alpha}E_{\vec k s \beta}^{(-)*} (\vec r', \omega),\label{eq:25}
  \\
  \vec k &= \frac{\omega}{c} \frac{\vec r}{r},\nonumber
\end{align}
where $\vec e_{s}$ are the polarization vectors and the fields
$E_{\vec ks \beta}^{(-)*} (\vec r', \omega)$ satisfy the homogeneous
Maxwell's equations
\begin{align}
  (\rot\rot \vec E_{\vec ks}^{(-)*}&(\vec r,\omega))_{\alpha}\label{eq:26}
  \\
  &-\frac{\omega^{2}}{c^{2}}\int d\vec r_{1}\epsilon_{\alpha\beta}^{*}(\vec r,\vec r_{1},\omega)E_{\vec ks\beta}^{(-)*}(\vec r_{1},\omega) = 0.\nonumber
\end{align}

It is important to stress here that the electromagnetic field
$\vec E_{\vec ks}$ emitted by the particle inside a crystal possesses
an asymptotic behavior for $r \to \infty$ of a plane wave and an
ingoing spherical wave
\begin{align}
  \vec E_{\vec k s}^{(-)*} (\vec r,\omega) \approx \vec
e_{s} \mathrm{e}^{\ri\vec k\cdot\vec r} + \vec f_{s}
\frac{\mathrm{e}^{-\ri k r}}{r}, \quad r \rightarrow \infty.\label{eq:27}
\end{align}
Here $\vec f_{s}$ are the scattering amplitudes, which are independent
of $r$.

In contrast, when an external electromagnetic field is scattered or
diffracted on a crystal, this wave possesses an asymptotic behavior of
outgoing spherical wave, which is usually denoted as
$\vec E_{\vec k s}^{(+)} (\vec r,\omega)$. However, the wave
$\vec E_{\vec k s}^{(-)*} (\vec r,\omega)$ and the
wave $\vec E_{\vec k s}^{(+)} (\vec r,\omega)$ are related to each
other with the following formula
\begin{align}
  (\vec E_{\vec k s}^{(-)} (\vec r,\omega))^{*} =  \vec E_{- \vec k s}^{(+)},\label{eq:28}
\end{align}
which is the analog of the well known reciprocity theorem in classical
optics \cite{born2013principles}.

Proceeding, we want to calculate the differential number of photons,
emitted in the solid angle $d\Omega$ and the spectral interval
$\omega,\omega+d\omega$. For this we consider that a particle of a
charge $e_{0}$ with the law of motion $\vec r(t)$ and the velocity
$\vec v(t) = d\vec r/dt$ generates a current, to be plugged in into
Eq.~(\ref{eq:20}). Then we take the asymptotic expression for the Green
function Eq.~(\ref{eq:25}) and substitute it into Eq.~(\ref{eq:24}),
which yields the electromagnetic field radiated in the far zone. This
field is then used in the expression for the energy density
\begin{align}
  W_{\vec n \omega} = \frac{c r^2}{4 \pi^2}|\vec E (\vec r,\omega)|^2. \label{eq:29}
\end{align}
Here $\vec n$ is the unit vector directed towards the observation
point $\vec n = \vec r/r$.

As a result the expression for the number of photons emitted in the
solid angle $d\Omega$ and the frequency range $\omega,\omega+d\omega$
can be obtained \cite{PXR_Book_Feranchuk}
\begin{align}
  \frac{\partial^2 N_{\vec n,\omega s}}{\partial \omega \partial
  \Omega} = \frac{e_0^2 \omega }{4 \pi^2 \hbar c^3} \left|\int \vec
  E_{\vec k' s}^{(-)\ast} (\vec r (t),\omega) \vec v (t) \ee^{\ri\omega t}
  dt\right|^2,\label{eq:30}
\end{align}
where $\vec k' = k \vec n$.

With the help of the reciprocity theorem we can transform
Eq.~(\ref{eq:30}) into the form
\begin{align}
  \frac{\partial^2 N_{\vec n,\omega s}}{\partial \omega \partial
  \Omega} &= \frac{e_0^2 \omega }{4 \pi^2 \hbar c^3} \left|\int \vec
            E_{\vec k s}^{(+)} (\vec r (t),\omega) \vec v (t) \ee^{\ri\omega t}
            dt\right|^2,\label{eq:31}
  \\
  \vec k &= - \vec k',\nonumber
\end{align}
where we have relabelled $-\vec k'$ with $\vec k$.

Let us apply this approach and calculate the emitted number of quanta
in the case of PXR-SPG. According to Eq.~(\ref{eq:31}) we first need to
solve a diffraction problem and determine the electromagnetic field
$\vec E_{\vec k s}^{(+)}$. For this we will apply the standard
approach, namely the two-wave approximation of the dynamical
diffraction theory \cite{authier2001dynamical,
  benediktovich2013theoretical}. In this case two strong
electromagnetic waves are excited in a crystal. The amplitudes of
these waves satisfy a set of algebraic equations
\cite{PXR_Book_Feranchuk}
\begin{align}
  \begin{aligned}
    \left(\frac{k^2}{k_0 ^2} -1 - \chi_0 \right) E_{\vec k s}  - c_s
    \chi_{-\vec g} E_{\vec k_g s} &= 0,
    \\
    \left(\frac{k_g^2}{k_0^2} -1 - \chi_0 \right) E_{\vec k_g s}  - c_s
    \chi_{\vec g} E_{\vec k s} &= 0.\label{eq:32}
  \end{aligned}
\end{align}
\begin{figure*}[t]
  \centering
  \includegraphics[width=0.48\textwidth]{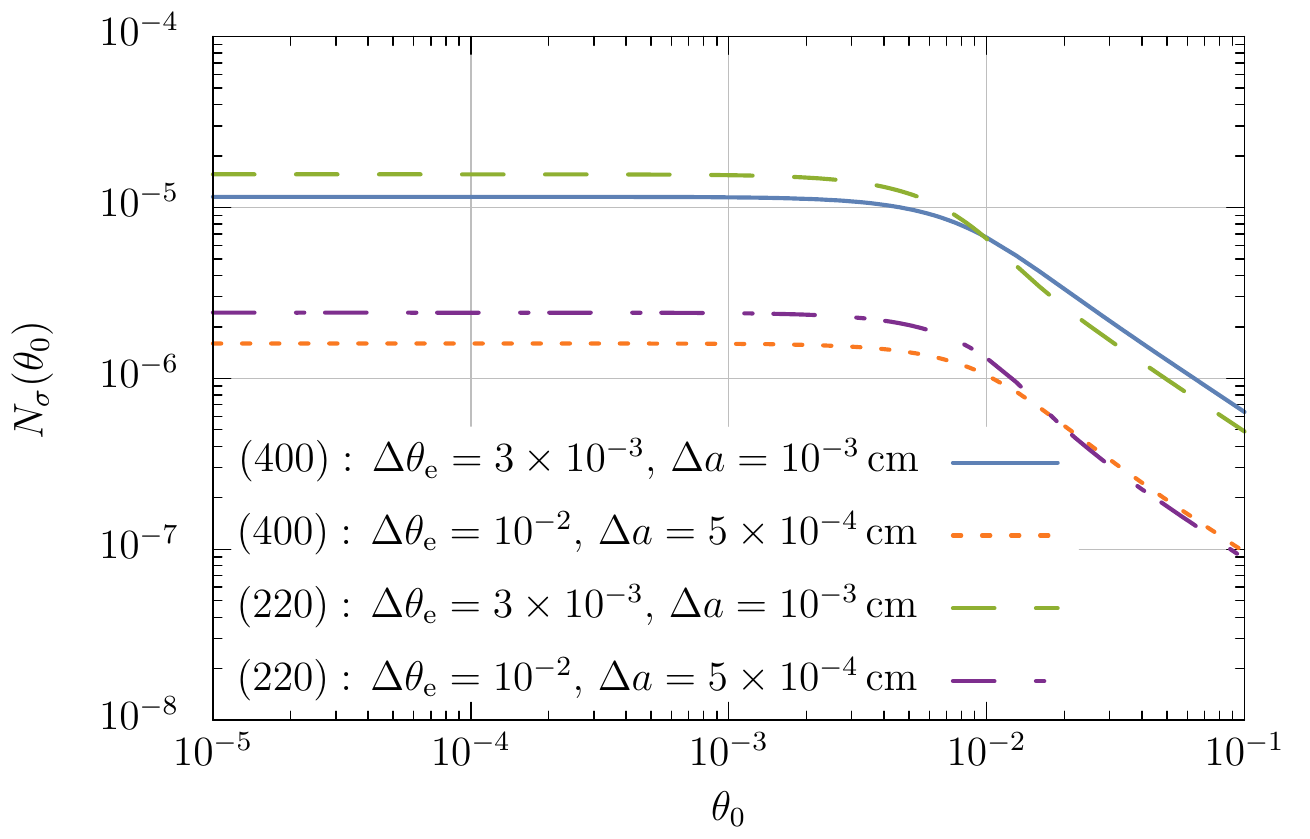}
  \includegraphics[width=0.48\textwidth]{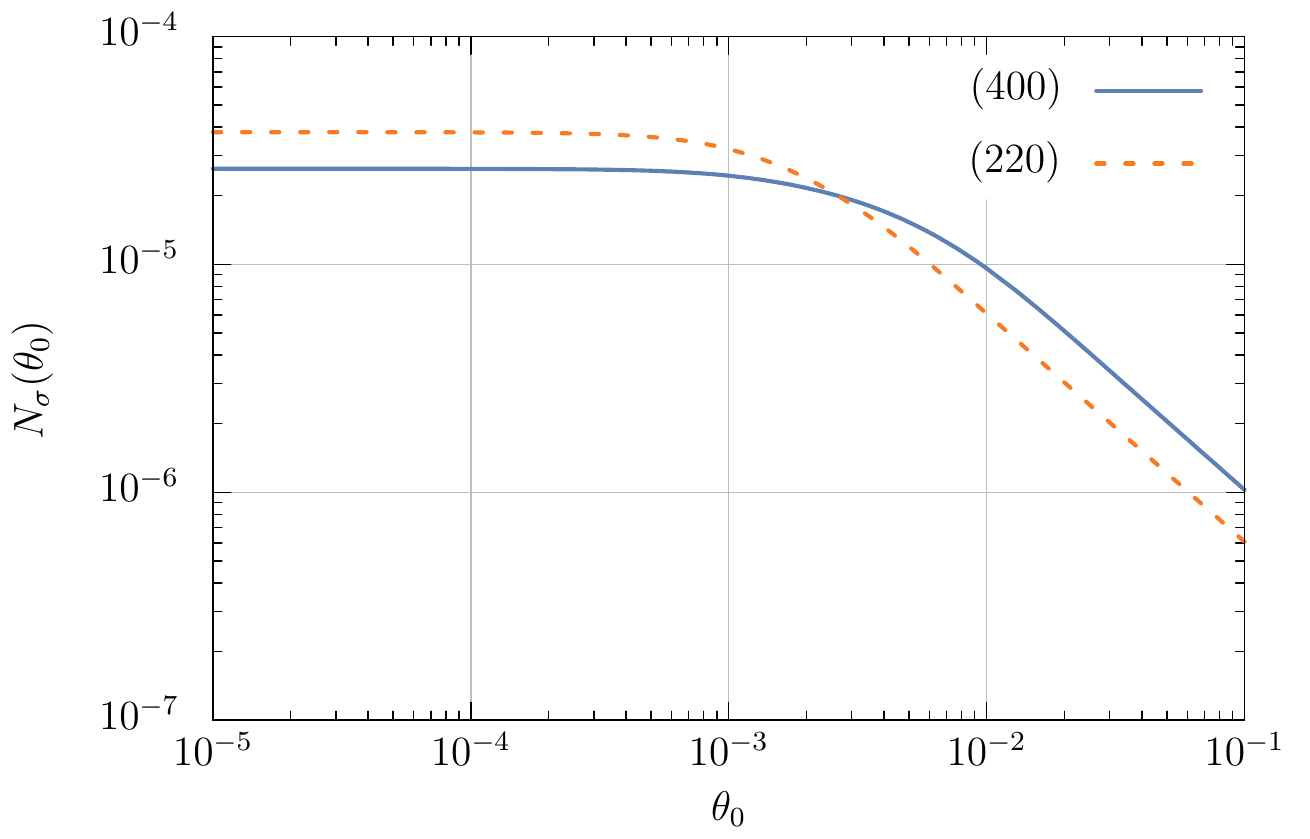}
  \caption{(Color online) Left pane: The dependence of the emitted
    number of quanta $N_{\mathrm{PXR-SPG}}$ on the incidence angle
    $\theta_{0}$ for the MAMI facility. The electron energy
    $E = 900\,\mathrm{MeV}$, the crystal length $L = 1\,\mathrm{cm}$
    and the distance from the center of the beam to the crystal
    surface $a_{0} = 0.5\times 10^{-3}\,\mathrm{cm}$. Right pane: The
    same dependence as on the left pane but for the LCLS facility. The
    electron energy $E = 8000\,\mathrm{MeV}$, the crystal length
    $L = 1\,\mathrm{cm}$, the electron beam spread
    $\Delta a = 5\times 10^{-3}\,\mathrm{cm}$ and the distance from
    the center of the beam to the crystal surface
    $a_{0} = 2\Delta a$.}\label{fig:3}
\end{figure*}
Here we introduced the scalar field amplitudes of the incident
$\vec E_{\vec k s}^{(+)} = \vec e_s E_{\vec k s}$ and diffracted
$\vec E_{\vec k_g s}^{(+)} =\vec e_{1s} E_{\vec k_g s}$ waves
respectively. In addition, $k_{0} = \omega/c$,
$\vec k_{g} = \vec k + \vec g$, $\vec g$ is the reciprocal
lattice vector, $\chi_{0}$ and $\chi_{\vec g}$ are the Fourier
components of the crystal susceptibility $\chi(\vec r)$
\begin{align}
   \chi(\vec r)= \sum_{\vec g} \chi_{\vec g} \ee^{\ri\vec g\cdot\vec
  r}.\label{eq:33}
\end{align}
The coefficient $c_s = 1 $ for the $\sigma$ polarization ($s=1$) and
$c_s = \cos 2\theta_B$ for the $\pi$ polarization ($s=2$) of the
incident and diffracted waves respectively.

For the following we note that the waves of different polarizations
propagate independently if we neglect terms of the order of
$\sim |\chi_{0}|^{2}$ in the Maxwell equations
\cite{authier2001dynamical, benediktovich2013theoretical}.

The system of equations (\ref{eq:32}) is a system of homogeneous linear
equations. Consequently, in order it to be solvable its determinant
should vanish. This provides us the dispersion equation for $k$ and
determines the relations between amplitudes of incident and diffracted
waves $E_{\vec ks}$ and $E_{\vec k_{g}s}$
\begin{align}
  \left(\frac{k^2}{k_0 ^2} -1 - \chi_0 \right)
  \left(\frac{k_g^2}{k_0^2} -1 - \chi_0 \right) - c_s^2\chi_{\vec g}
  \chi_{-\vec g} = 0, \label{eq:34}
  \\
  E_{\vec k_g s} = V_{\vec k s} E_{\vec k s}, \quad V_{\vec k s} =
  \frac{(\frac{k^2}{k_0^2} -1 - \chi_0 )  }{c_s \chi_{-\vec g}
  }.\label{eq:35}
\end{align}

In Fig.~\ref{fig:2} we show the propagation directions of
electromagnetic waves in vacuum and in crystal respectively. We remind
here that in the X-ray diffraction theory this type of geometry
corresponds to the EAD case, namely grazing exit
\cite{authier2001dynamical, benediktovich2013theoretical}. As follows
from Fig.~\ref{fig:2} that in order to satisfy the boundary conditions
on the interface one needs to take into account in vacuum not only an
incident wave $\vec E^{(0)}_{\vec ks}$, but also a specularly
reflected diffracted wave $\vec E^{(sp)}_{\vec k_{g}s}$, which defines
the PXR-SPG
\begin{align}
  \vec E_{\vec k s}^{(0)} &= \vec e_s \ee^{\ri \vec k\cdot\vec r}, \label{eq:36}
  \\
  \vec E_{\vec k_g s}^{(sp)} &= \vec e_{1s} E_{\vec k_g s}^{(sp)} \ee^{\ri
  (\vec k_{\|} + \vec g_{\|})\cdot\vec r} \ee^{\ri  k'_{gz}  z}, \label{eq:37}
  \\
  k'_{gz} &=  \sqrt{k_0^2 - (\vec k_{\|} + \vec g_{\|})^2 } \nonumber
\end{align}

The dispersion equation (\ref{eq:34}) is a fourth-order algebraic
equation and consequently it has four solutions. For this reason there
exist four electromagnetic waves in a crystal. However, two out of four
these solutions are unphysical, since they lead to the exponentially
growing field amplitudes inside the crystal. Consequently, when a
boundary problem is solved only two waves should be taken into
account. These waves have positive imaginary part of the $k_{z}$, i.e,
the $z$-component of the wave vector inside a medium. According to the
boundary conditions the in-plane components $\vec k_{\|}$ of the wave
vector are conserved. Therefore, the change of the wave vector is
defined through the projection on the $\vec N$ - the normal to the
crystal surface \cite{PXR_Book_Feranchuk}
\begin{align}
  \vec k_{s \mu} = \vec k  - k_{0} \epsilon_{\mu s}\vec N, \quad \mu =
  1,2,\label{eq:38}
\end{align}
where $\vec k = k_{0}\vec n$ and $\vec n$ is the unit vector of the
incident wave in vacuum. At the same time, the wave vector $\vec
k_{g}$ of the diffracted wave is directed parallel to the crystal
surface and, therefore, the quantity $|\nu_{g}| = |\vec k_{g}\cdot\vec
N / k_{0}|$ is much smaller than unity, i.e., $|\nu_{g}|\ll 1$. For
the following we also define the quantity $\nu_{0} = \vec k\cdot\vec N/k_{0}$.
\begin{figure}[t]
  \centering
  \includegraphics[width=0.48\textwidth]{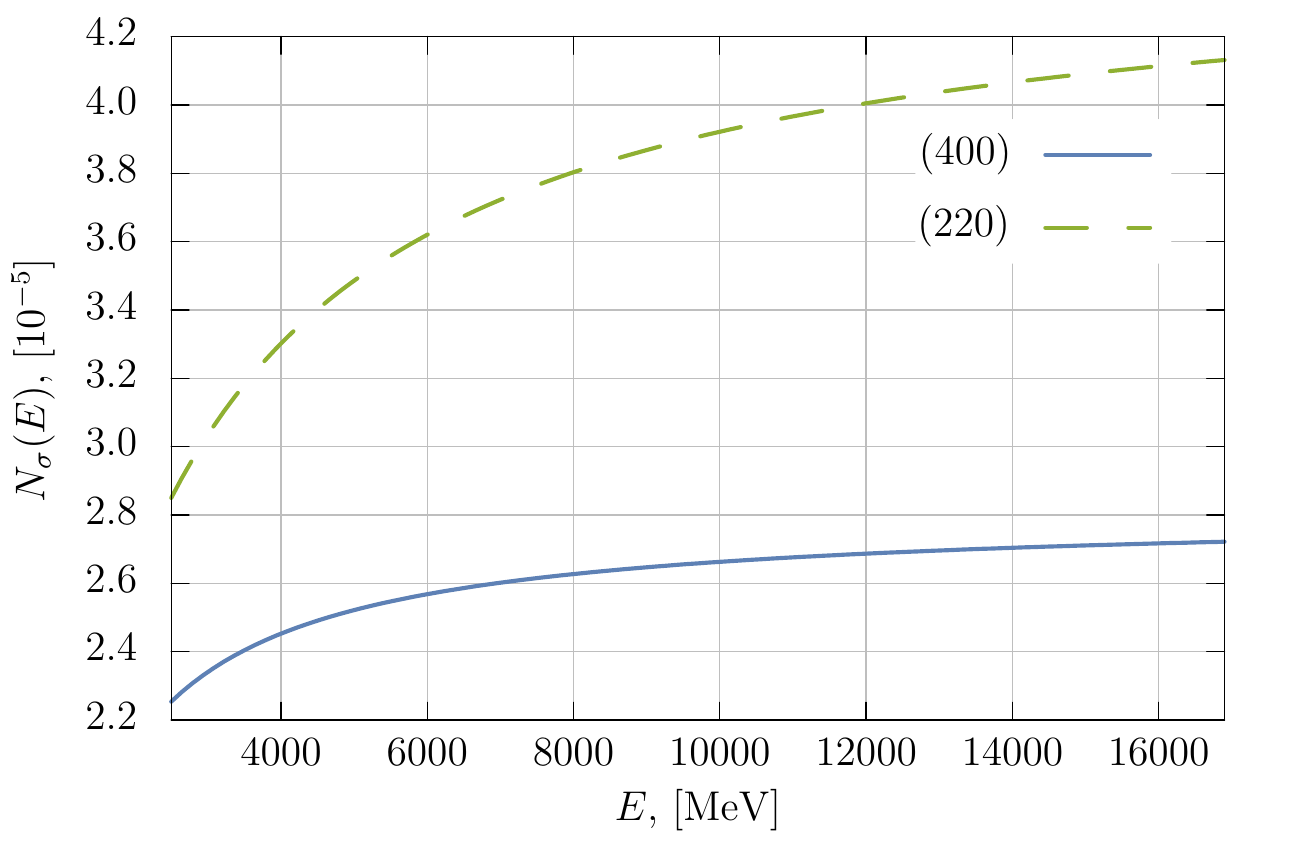}
  \caption{(Color online) The dependence of the total number of the
    emitted photons $N_{\mathrm{PXR-SPG}}$ on the electron beam energy
    for the LCLS facility. The parameters of the reflection are given
    via Eqs.~(\ref{eq:61}), (\ref{eq:62}). The electron beam
    transverse size $\Delta a = 5\times 10^{-3}\,\mathrm{cm}$, the
    distance from the beam center to the crystal $a_{0} = 2\Delta a$,
    the crystal length $L = 1\,\mathrm{cm}$ and the angle
    $\theta_{0} = 10^{-5}\,\mathrm{rad}$.}\label{fig:5}
\end{figure}

We now plug in the expression (\ref{eq:38}) into the dispersion
equation (\ref{eq:32}) and neglect the specularly reflected incident
wave, which amplitude is proportional to $|\chi_{0}|^{2}$
\cite{authier2001dynamical, benediktovich2013theoretical} in the
considered geometry. This yields the following cubic equation for the
values of $\epsilon_{s\mu}$
\begin{align}
  - 2 \nu_{0} \epsilon_{\mu s}^3 &+ (4\nu_{0}\nu_{g} - \chi_0)
                                    \epsilon_{\mu s}^2 +
  2 \nu_{0} (\chi_0 - \alpha_{\mathrm{B}})\epsilon_{\mu s}\label{eq:39}
  \\
  &\mspace{80mu}+ \chi_0^2 - \chi_0 \alpha_{\mathrm{B}} - c_s^2\chi_{\vec
    g}\chi_{-\vec g} = 0,
  \nonumber
  \\
  \alpha_{\mathrm{B}} &= \frac{k^2 - k_g^2}{k_0^2} = - \frac{2 \vec k\cdot\vec g +
  g^2}{k_0^2}.\nonumber
\end{align}
In this equation the parameter $\alpha_{\mathrm{B}}$ defines the
deviation from the Wulff-Bragg condition.

In the general case the solutions of this cubic equation are given
through the cumbersome Cardano formula. However, these solutions can
be significantly simplified if we employ the approximation, described
in the Ref.~\cite{PSSA:PSSA2210710211}. For this we note that the
angular spread of the PXR peak is given via a parameter
$\theta_{0}\simeq \sqrt{|\chi_{0}|}$. In addition, in the considered
geometry $|\nu_{0}|\approx 1$ and
$|\nu_{g}| \approx \sqrt{|\chi_{0}|}\gg |\chi_{0}|$. In this
approximation the desired solutions for the decaying fields inside a
crystal are simplified and are equal to
\begin{align}
  \epsilon_{1s} &= - \frac{\chi_0}{2 \nu_{0}} + \frac{c_s^2\chi_{\vec
  g}\chi_{-\vec g}}{2\nu_{0} (\alpha_{\mathrm{B}} + \chi_0) },\label{eq:40}
  \\
  \epsilon_{2s} &= \nu_{g}  + \sqrt{\nu_{g}^2 + \alpha_{\mathrm{B}} +
                  \chi_0}, \label{eq:41}
  \\
  |\epsilon_{2s}| &\sim \sqrt{|\chi_0|} \gg  |\epsilon_{1s}| \sim
  |\chi_0|, \label{eq:42}
  \\
  \epsilon_{2s}'' &\sim \frac{\chi_0''}{\sqrt{|\chi_0|}} \gg
                    \epsilon_{1s}'' \sim \chi_0''. \label{eq:43}
\end{align}

Having found solutions of the dispersion equation we can write down
the solution of the Maxwell's equation in vacuum $z > 0$ and inside a
crystal $z < 0$ as the following superposition
\begin{align}
  \vec E_{\vec k s}^{(+)}
  &= \vec e_s \ee^{\ri \vec k\cdot\vec r} + \vec e_{1s} E_{  s}^{(sp)}
    \ee^{\ri (\vec k_{\|} + \vec g_{\|})\cdot\vec r} \ee^{\ri k'_{gz}
    z}, \ z > 0, \label{eq:44} 
  \\
  \vec E_{\vec k s}^{(+)}
  &= \ee^{\ri \vec k\cdot\vec r}\sum_{\mu = 1,2}\ee^{-\ri k_0 z
    \epsilon_{\mu s}}(\vec e_s E_{\mu s} \nonumber
  \\
  &\mspace{180mu}+ \vec e_{1s} E_{ g \mu s}\ee^{\ri \vec{g}\cdot\vec
    r}), \ z < 0, \label{eq:45}
\end{align}
where the coordinate $x$ is changing in the limits $0 < x < L$, with
$L$ being the crystal length (see Fig.~\ref{fig:2}).

In order to determine the amplitudes of these waves we require the
continuity of the electromagnetic field on the crystal surface
\cite{authier2001dynamical, benediktovich2013theoretical}, which in
our case yields the system of equations
\begin{equation}
  \label{eq:46}
  \begin{aligned}
    &E_{1 s} + E_{2 s} = 1,
    \\
    &E_{g 1 s} + E_{g 2 s} = E_{s}^{(sp)},
    \\
    &(\nu_{g} - \epsilon_{1 s}) E_{g 1 s} + (\nu_{g} - \epsilon_{2 s})
    E_{g 2 s} = \nu_{g}' E_{s}^{(sp)},
  \end{aligned}
\end{equation}
where $\nu_{g}' = k'_{gz} / k_{0}$. The solution of this
system of equations can be easily obtained, which reads
\begin{align}
  E_{1 s}  &= - \frac{(\chi_0 + 2 \epsilon_{2 s} \nu_{0}) (\nu_{g}' -
  \nu_{g} +\epsilon_{2 s}) }{(\epsilon_{1 s} - \epsilon_{2
  s})[2\nu_{0} (\nu_{g}' - \nu_{g}  + \epsilon_{2 s} + \epsilon_{1
  s}) + \chi_0 ]}, \nonumber
  \\
  E_{2 s}  &= - \frac{(\chi_0 + 2 \epsilon_{1 s} \nu_{0}) (\nu_{g}' -
  \nu_{g} +\epsilon_{1 s}) }{(\epsilon_{1 s} - \epsilon_{2
  s})[2\nu_{0} (\nu_{g}' - \nu_{g}  + \epsilon_{2 s} + \epsilon_{1
  s}) + \chi_0 ]}, \nonumber
  \\
  E_{g 1 s}  &= - \frac{(\chi_0 + 2 \epsilon_{1 s} \nu_{0})(\chi_0 + 2
  \epsilon_{2 s} \nu_{0}) (\nu_{g}' - \nu_{g} +\epsilon_{2 s})
  }{c_s \chi_{-g}(\epsilon_{1 s} - \epsilon_{2 s})[2\nu_{0}
  (\nu_{g}' - \nu_{g}  + \epsilon_{2 s} + \epsilon_{1 s}) + \chi_0
  ]}, \nonumber
  \\
  E_{g 2 s}  &= - \frac{(\chi_0 + 2 \epsilon_{1 s} \nu_{0})(\chi_0 + 2
  \epsilon_{2 s} \nu_{0}) (\nu_{g}' - \nu_{g} +\epsilon_{1 s})
  }{c_s \chi_{-g}(\epsilon_{1 s} - \epsilon_{2 s})[2\nu_{0}
  (\nu_{g}' - \nu_{g}  + \epsilon_{2 s} + \epsilon_{1 s}) + \chi_0
  ]}, \nonumber
  \\
  E^{(sp)}_s   &= - \frac{(\chi_0 + 2 \epsilon_{1 s} \nu_{0})(\chi_0 +
  2 \epsilon_{2 s} \nu_{0})   }{c_s \chi_{-g} [2\nu_{0} (\nu_{g}' -
  \nu_{g}  + \epsilon_{2 s} + \epsilon_{1 s}) + \chi_0 ]}.\label{eq:47}
\end{align}

According to its definition, the PXR is emitted by a particles, which
move uniformly. This corresponds to the following law of motion
\begin{align}
  \vec r(t) = \vec r_0 + \vec v t, \quad  \vec r_0 = (0,0,z_0),\label{eq:48}
\end{align}
where $\vec r_{0}$ is the coordinate of an electron in the $(x,z)$
plane, which is perpendicular to the crystal surface. In addition, we
suppose that the $r_{0y} = 0$ and in the moment of time $t=0$ the $x$
coordinate of a particle is equal to zero, i.e., $x = 0$.

When $\vec v \bot \vec N$ the radiation is formed by the field of the
diffracted wave, which propagates in the direction, defined via an
electron velocity $\vec v$ that is parallel to the vector
$\vec k_{g} = \vec k + \vec g$, i.e., $\vec v \| \vec k_{g}$. For this
wave the Cherenkov condition $\vec v\cdot\vec k_{g} = \omega/c$ can be
fulfilled \cite{J.Phys.France1983.44.913}. In our case the particle
moves in vacuum, parallel to the crystal vacuum interface, which
corresponds to $z_{0} < 0$. Consequently, the radiation is defined by
the specularly reflected diffracted wave $E_{s}^{sp}$.
\begin{figure*}[t]
  \centering
  \includegraphics[width=0.48\textwidth]{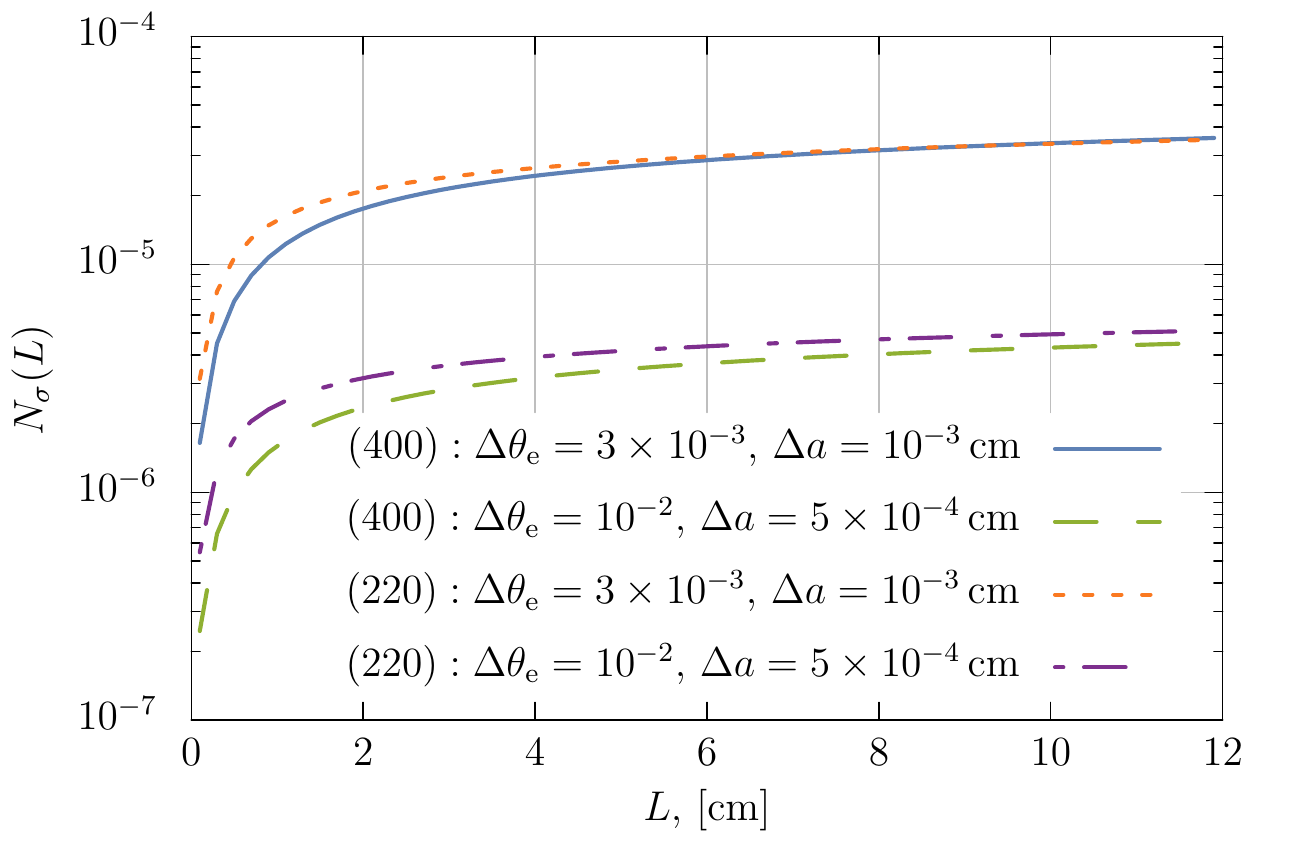}
  \includegraphics[width=0.48\textwidth]{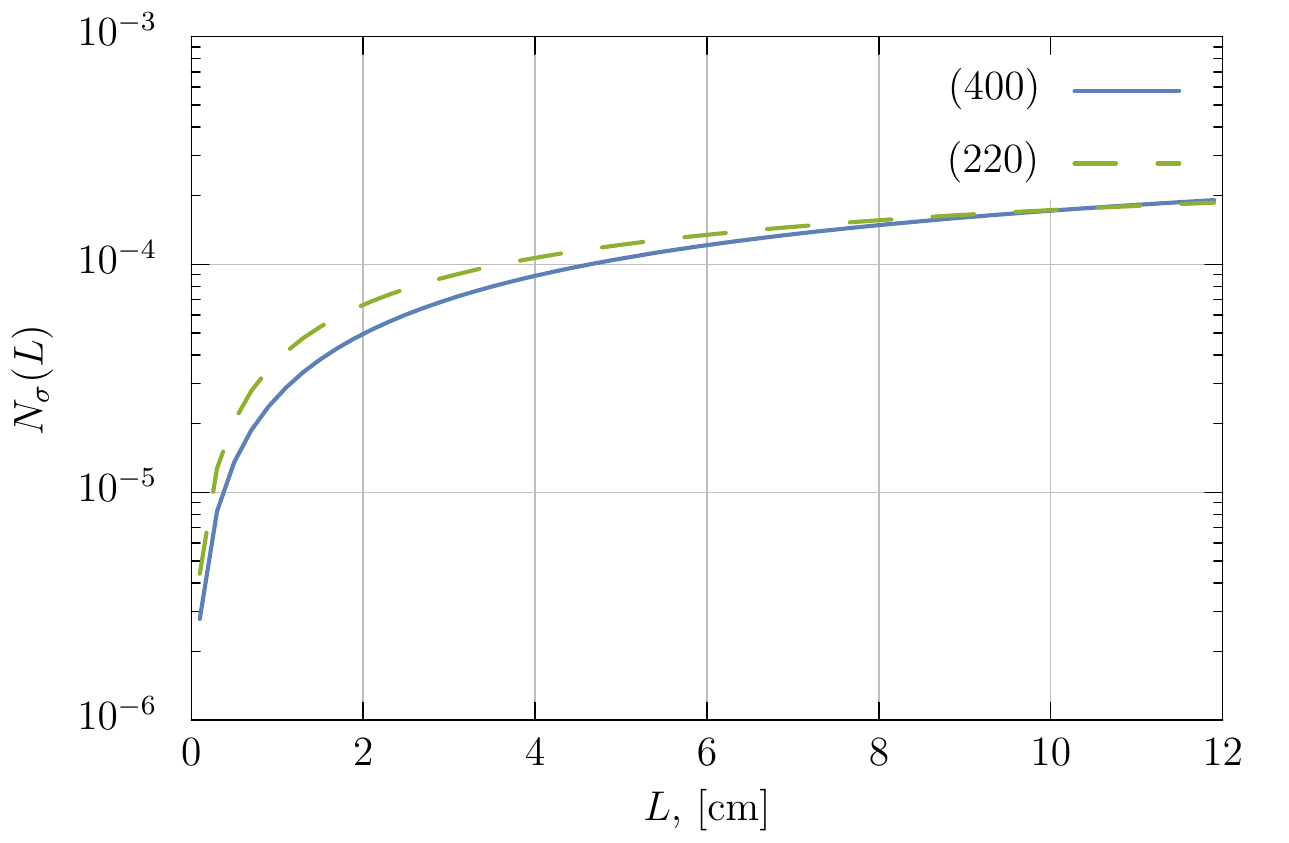}
  \caption{(Color online) Left pane: The dependence of the emitted
    number of quanta $N_{\mathrm{PXR-SPG}}$ on the crystal length $L$
    for the MAMI facility. The electron energy
    $E = 900\,\mathrm{MeV}$, angle
    $\theta_{0} = 10^{-5}\,\mathrm{rad}$ and the distance from the
    center of the beam to the crystal surface
    $a_{0} = 0.5\times 10^{-3}\,\mathrm{cm}$. Right pane: The same
    dependence as on the left pane but for the LCLS facility. The
    electron energy $E = 8000\,\mathrm{MeV}$, angle
    $\theta_{0} = 10^{-5}\,\mathrm{rad}$, the electron beam spread
    $\Delta a = 5\times 10^{-3}\,\mathrm{cm}$ and the distance from
    the center of the beam to the crystal surface $a_{0} = 2\Delta
    a$. In addition we assume that the electrons of a beam are
    described via distribution function (\ref{eq:60}) with the mean
    values of $\theta_{0}$ and $a_{0}$ respectively.}\label{fig:6}
\end{figure*}

Continuing, we plug in the law of motion (\ref{eq:48}) into
Eq.~(\ref{eq:31}) for the differential number of photons and integrate
over the particle trajectory. This yields
\begin{align}
  \frac{\partial^2 N_{\vec n,\omega s}}{\partial \omega  \partial
  \Omega} &= \frac{e_0^2  \omega }{4 \pi^2 \hbar c^5} (\vec e_{1s}\cdot\vec
  v)^2 \nonumber
  \\
  &\times \left|E^{(sp)}_s   L_g (1 - \ee^{- \ri L/L_g}) \ee^{\ri k_{gz}' z_0
    }\right|^2,\label{eq:49}
\end{align}
where
\begin{align}
  \begin{aligned}
    L_g &= \frac{c}{\omega - v_x(k_x + g_x) -  v_z k_{gz}' } \equiv
    \frac{1}{k_0 q},
    \\
    q &= 1 - \frac{ v_x(k_x + g_x)}{\omega} - \frac{v_z k_{gz}'}{\omega},
    \\
	k_{gz}' &= \sqrt{k_0^2 - (k_x + g_x)^2 - k_y^2 }, \quad g_{y} = 0.
  \end{aligned}\label{eq:50}
\end{align}

Here the quantity $L_{g}$ determines the coherence length
\cite{Galitsky1964} for the radiation mechanism under investigation
and $k_{0}q$ is the longitudinal component of the wave vector, which
was transferred to the electron during the emission of a photon
\cite{J.Phys.France1983.44.913}
($kq \approx (p_{z} - p_{1z}) / \hbar - k_{z}$ with $p_{z}$ and
$p_{1z}$ being the electron momenta before and after radiation of a
photon correspondingly). The intensity of radiation has the maximal
value for frequencies and angles of the emitted photons when the
quantity $|q^{2}|$ is minimal. In the considered geometry the
following condition is fulfilled
\begin{align}
  v_x \approx v(1 - \theta_{\mathrm{e}}^2/2) \gg v_z \approx v
  \theta_{\mathrm{e}}, \label{eq:51}
\end{align}
where the angle $\theta_{\mathrm{e}}$ is the angle between the
electron velocity and the crystal surface (see
Fig.~\ref{fig:2}). Therefore, as follows from Eq.~(\ref{eq:50}) the
maximum of the coherence length is achieved when
\begin{align}
  q_0 &= 1 - \frac{ v_x(k_x + g_x)}{\omega} \approx 0,\nonumber
  \\
  \omega &\approx \omega_{\mathrm{B}} (1 - \theta_{\mathrm{e}}^2/2 +
  \theta_x), \quad \omega_{\mathrm{B}}  = v g_{x},\label{eq:52}
\end{align}
which defines the frequency of the PXR-SPG. Here $\theta_{x}$ is the
angle between the wave vector of the emitted photon and the normal
$\vec N$ to the crystal surface. We also note here
that the frequency of the emitted radiation coincides with the
frequency of the Smith-Purcell effect, when the period of the
surface grating is equal to $d = 2\pi / g_{x}$.
\begin{figure*}[t]
  \centering
  \includegraphics[width=0.48\textwidth]{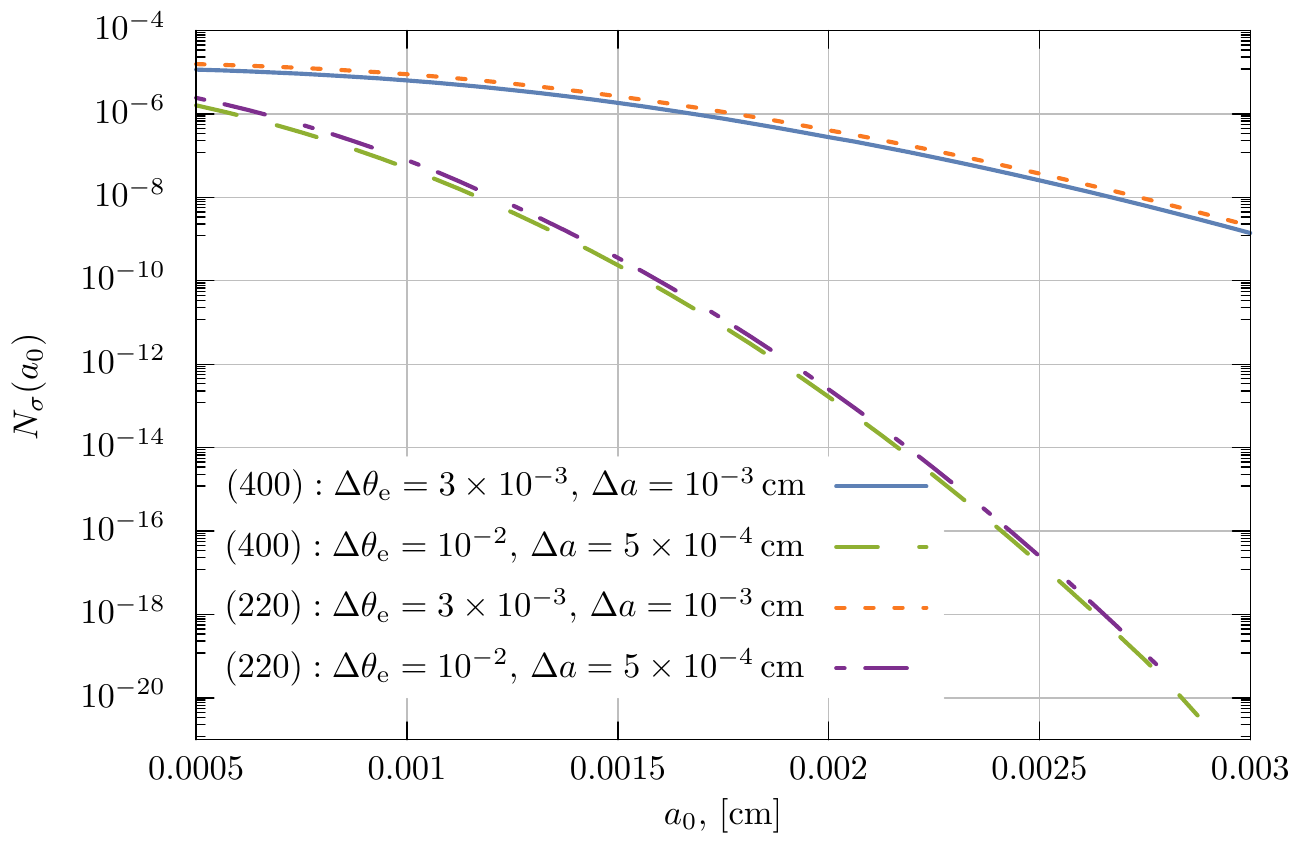}
  \includegraphics[width=0.48\textwidth]{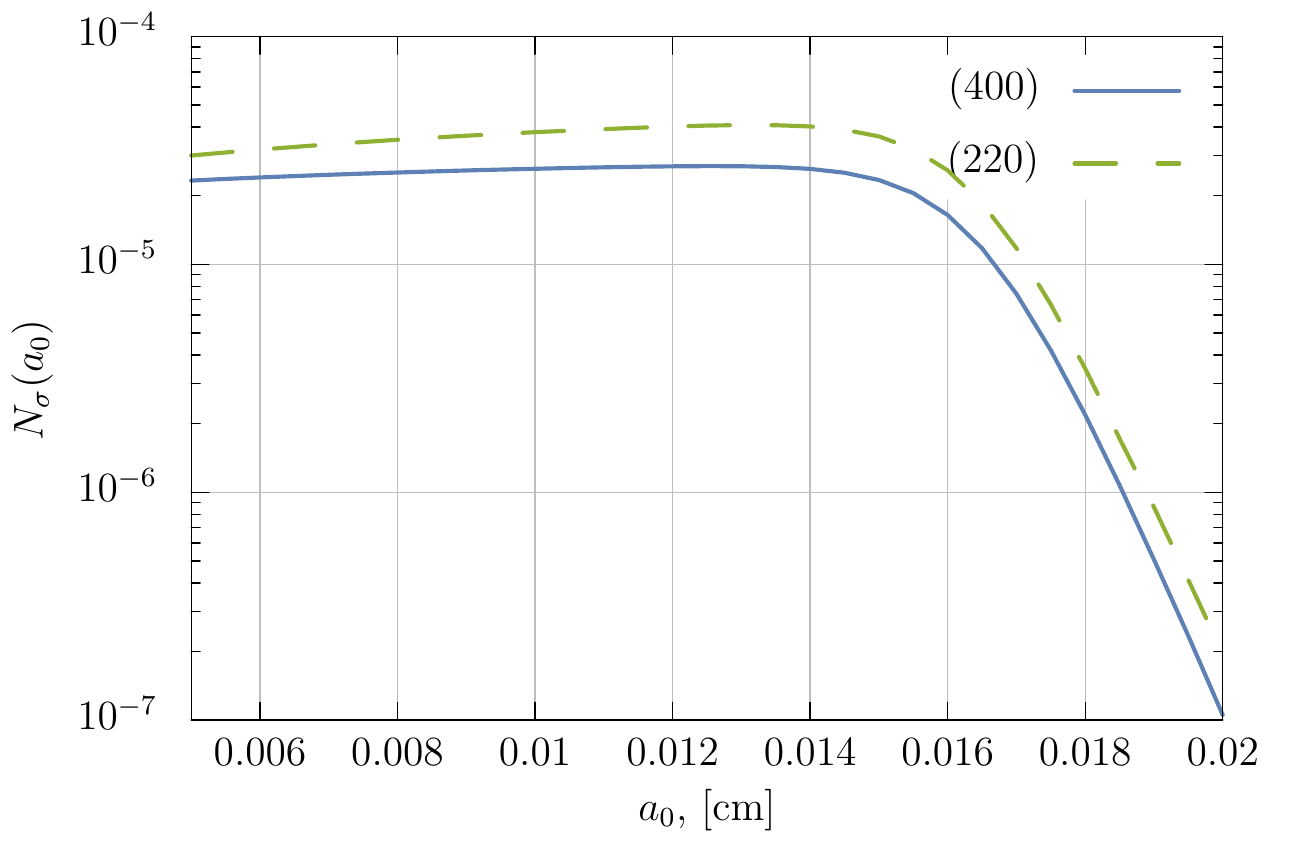}
  \caption{(Color online) Left pane: The dependence of the emitted
    number of quanta $N_{\mathrm{PXR-SPG}}$ on the distance from the
    beam center to the crystal surface $a_{0}$ for the MAMI
    facility. The electron energy $E = 900\,\mathrm{MeV}$, angle
    $\theta_{0} = 10^{-5}\,\mathrm{rad}$ and the crystal length
    $L = 1\,\mathrm{cm}$. Right pane: The same dependence as on the
    left pane but for the LCLS facility. The electron energy
    $E = 8000\,\mathrm{MeV}$, angle
    $\theta_{0} = 10^{-5}\,\mathrm{rad}$, the electron beam spread
    $\Delta a = 5\times 10^{-3}\,\mathrm{cm}$ and the crystal length
    $L = 1\,\mathrm{cm}$.}\label{fig:7}
\end{figure*}

In this frequency range the $z$-component of the wave vector $k'_{gz}$
is purely imaginary, such that
\begin{align}
  k_{gz}' &= \sqrt{k_0^2 - \frac{\omega^{2}}{v_x^2} - k_{y}^{2}}\nonumber
  \\
  &\approx \ri k_0 \sqrt{(\theta_x -\theta_{\mathrm{e}z})^2 + \theta_y^2
  +\theta_{\mathrm{e}y}^2 + \gamma^{-2}} \equiv \ri k_{0} \eta, \label{eq:53}
\end{align}
Consequently, the phase of the exponent $\exp(\ri k'_{gz}z_{0})$ in
the Eq.~(\ref{eq:49}) becomes real and the intensity of radiation
exponentially decays with the increase of the distance from the
crystal surface.

The field amplitude $E_{s}^{(sp)}$ to be inserted in Eq.~(\ref{eq:49})
significantly simplifies under the assumptions (\ref{eq:42}),
(\ref{eq:43}) and looks like
\begin{align}
  E^{(sp)}_s = \frac{c_s \chi_g }{\alpha_{\mathrm{B}} + \chi_0}. \label{eq:54}
\end{align}

In the Smith-Purcell geometry the angle $\theta_{\mathrm{B}} = \pi/4$,
$c_{\sigma} = 1$, $c_{\pi} = \cos2\theta_{\mathrm{B}} \approx 0$. This
means that the emitted radiation will be polarized in the direction
perpendicular to the plane defined by the vectors $\vec N$ and $\vec
g$. In addition, in the point where the intensity of radiation is
maximal, which corresponds to $q_{0} = 0$, the quantity
$\alpha_{\mathrm{B}}$ exactly coincides with the one in the case of
PXR-EAD \cite{SKOROMNIK201786}. That is
\begin{align}
  \alpha_{\mathrm{B}} &= -2[ 1 - v_0 \cos \theta \cos
  \theta_{\mathrm{e}} - \theta_y \theta_{\mathrm{e}y} - \theta_x
  \theta_{\mathrm{e}z}] \nonumber
  \\
  &\approx -[ \gamma^{-2} + (\theta_y - \theta_{\mathrm{e}y})^2 +
    (\theta_x -\theta_{\mathrm{e}z})^2 ]. \label{eq:55}
\end{align}

Combining together all expressions and plugging them into
Eq.~(\ref{eq:49}) we determine the spectral and angular distribution
for the number of photons emitted per electron in the frequency range
$\omega,\omega + d\omega$, in the solid angle $d\Omega\approx
d\theta_{x}d\theta_{y}$
\begin{align}
  \frac{\partial^3 N_{ \sigma}}{\partial \omega \partial \theta_x
  \partial \theta_y} &= \frac{e_0^2}{4 \pi^2 \hbar\omega  c }
  \frac{(\theta_y - \theta_{\mathrm{e}y})^2 |\chi_g|^2 }{[\gamma^{-2} +
  (\theta_y - \theta_{\mathrm{e}y})^2 + (\theta_x -\theta_{\mathrm{e}z})^2 -
  \chi_0']^2} \nonumber
  \\
  &\times  \frac{[1 - 2\cos(k_{0}q_{0}L)\ee^{-
    k_0 L \theta_{\mathrm{e}z} \eta} + \ee^{- 2 k_0 L \theta_{ez}\eta}
    ]}{q_0^2 + \theta_{\mathrm{e}z}^2 \eta^2}\nonumber
  \\
  &\times \ee^{ -2 k_0 \eta |z_0|}\label{eq:56}
\end{align}
and all parameters of the medium are evaluated at the frequency
$\omega = \omega_{\mathrm{B}}$.

The integration with respect to frequency can be performed in the
complex plane of the variable $q_{0}$, taking into account that
$dq_{0} = d\omega / \omega_{\mathrm{B}}$. This yields
\begin{align}
  \frac{\partial^2 N_{ \sigma}}{\partial \theta_x  \partial
  \theta_y} &= \frac{e_0^2}{4 \pi  \hbar  c } \frac{(\theta_y -
  \theta_{\mathrm{e}y})^2 |\chi_g|^2 }{[\gamma^{-2} + (\theta_y -
  \theta_{\mathrm{e}y})^2 + (\theta_x -\theta_{\mathrm{e}z})^2 - \chi_0']^2}
  \nonumber
  \\
  &\times \frac{[ 1 - \ee^{- 2 k_0 L |\theta_{ez}|\eta}
    ]}{|\theta_{\mathrm{e}z}| \eta} \ee^{-2 k_0 \eta |z_0|}\label{eq:57}
\end{align}
This equation defines the intensity of the emitted radiation from a
part of the beam, which moves in vacuum $z_{0} > 0$, parallel to
the crystal vacuum interface, where the parameter $\eta$ is defined as
\begin{align}
  \eta &= \sqrt{(\theta_x -\theta_{\mathrm{e}z})^2 + \theta_y^2
  +\theta_{\mathrm{e}y}^2 + \gamma^{-2}}. \label{eq:58}
\end{align}

According to Eq.~(\ref{eq:3}) this electrons contribute to the
Smith-Purcell intensity $N_{\mathrm{SP}}$.

In contrast, electrons of a beam that are moving inside the crystal,
i.e., $z_{0} < 0$ emit radiation in the geometry of PXR-EAD,
which was recently investigated in Ref.~\cite{SKOROMNIK201786}. The
differential number of photons emitted in this case reads
\begin{align}
  \frac{\partial^2 N_{ \sigma}}{\partial \theta_x  \partial \theta_y}
  &= \frac{e_0^2}{4 \pi \hbar c } \frac{(\theta_y -
    \theta_{\mathrm{e}y})^2 |\chi_g|^2 }{[\gamma^{-2} + (\theta_y -
    \theta_{\mathrm{e}y})^2 + (\theta_x -\theta_{\mathrm{e}z})^2 -
    \chi_0']^2}\times
    \nonumber
  \\
  &\times  \frac{(1 - \ee^{-L k_0\chi_0''|\theta_{\mathrm{e}z}|}
    )}{\chi_0''|\theta_{\mathrm{e}z}|} \ee^{ -\chi_0'' k_0 |z_0| }.
    \label{eq:59}
\end{align}
We notice here that the spectral angular distribution of the emitted
photons depends on the parameters of the crystalline planes only
through the components of dielectric susceptibilities.

In order to evaluate the total number of photons emitted by the bunch
of electrons, one needs to integrate Eqs. (\ref{eq:57}), (\ref{eq:59})
over angles and average over the parameters of the electron beam. This
is performed by making a convolution with the distribution functions
over the initial position $z_{0}$ and the spreads of the electron
angles $\theta_{\mathrm{e}y,z}$ due to the bunch emittance when
$z_{0} > 0$ and, in addition, due to the multiple electron scattering
when $z_{0} < 0$. We consider the Gaussian distribution functions over
the electron beam parameters that are given as
\begin{align}
  G
  &(\theta_{\mathrm{e}z}, \theta_{\mathrm{e}y}, z_0) = C
    F(\theta_{\mathrm{e}z}, \theta_{\mathrm{e}y}, z_0),
    \nonumber
  \\
  F
  &=  \exp[-((\theta_{\mathrm{e}z}-\theta_{0})^{2} + \theta_{\mathrm{e}y}^{2}) /
    \Delta\theta_{\mathrm{e}}^{2} ]
    \label{eq:60}
  \\
  &\mspace{40mu}\times \exp[- ( z_{0}  - a_{0})^{2} / \Delta a^{2}],
    \quad z_0 > 0
    \nonumber
  \\
  F &=  \exp[-(\theta_{\mathrm{e}z}^{2} + \theta_{\mathrm{e}y}^{2}) /
      (\theta_{\mathrm{s}}^{2} + \Delta\theta_{\mathrm{e}}^{2})]
      \nonumber
  \\
  &\mspace{40mu}\times \exp[- ( z_{0}  - a_{0})^{2} / \Delta a^{2}], \quad
    z_0 < 0.\nonumber
\end{align}
Here the angle $\theta_{0}$ is the incident angle of the electron beam
and $\Delta a$ and $\Delta\theta_{\mathrm{e}}$ are the transversal and
angular spreads correspondingly. When the angle $\theta_{0}$ is close
to zero we have the grazing geometry of PXR-EAD and conversely, when
it is large on has the conventional transition geometry of PXR. The
constant $C$ determines the normalization condition of the
distribution. We choose it from requiring that the distribution is
normalized to one
\begin{align*}
  \int d \theta_{\mathrm{e}z} d \theta_{\mathrm{e}y} d z_0
  G(\theta_{\mathrm{e}z}, \theta_{\mathrm{e}y}, z_0) = 1,
\end{align*}
such that the total intensity is referred to the single electron,
i.e., the total intensity from the electron bunch is divided by the
number of electrons in the bunch.

Tanking into account the above discussion, for the numerical
evaluation we consider that the angle $\theta_{\mathrm{e}z}$ is
counted from the angle $\theta_{0}$ and introduce the polar
coordinates $\theta_{x} - \theta_{0} = \rho\cos\varphi$,
$\theta_{y} = \rho\sin\varphi$. The integration is then performed in
the range $\rho = [0, \theta_{\mathrm{D}}]$ and $\varphi = [0,2\pi]$,
with $\theta_{\mathrm{D}}$ being the aperture of a detector. The
result of this integration is then convoluted with the distribution
function given via Eq.~(\ref{eq:60}).

\section{Numerical results and discussion}
\label{sec:numer-results-disc}
\begin{figure*}[t]
  \centering
  \includegraphics[width=0.48\textwidth]{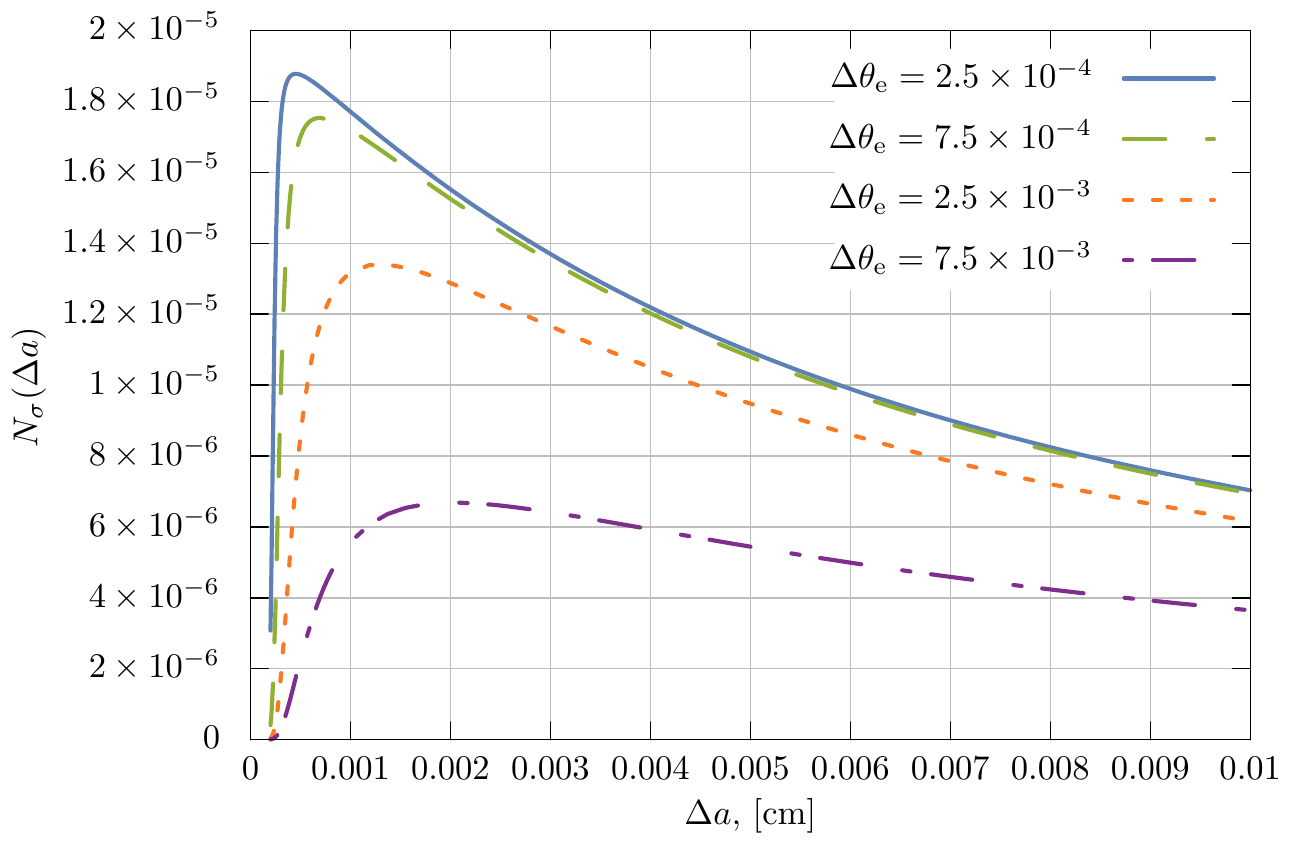}
  \includegraphics[width=0.48\textwidth]{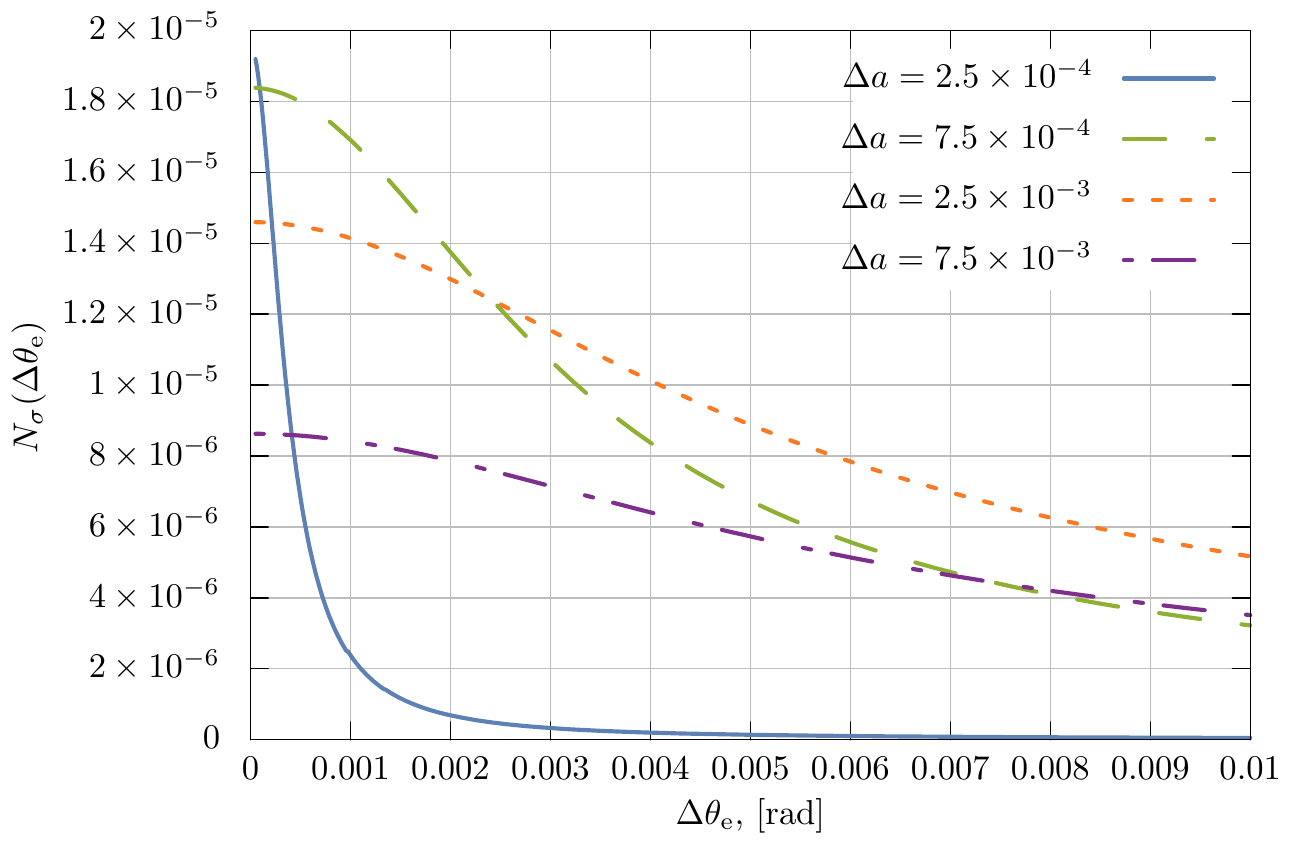}
  \caption{(Color online) Left pane: The dependence of the emitted
    number of quanta on the beam transversal size $\Delta a$ for the
    MAMI facility. The electron energy $E = 900\,\mathrm{MeV}$, angle
    $\theta_{0} = 10^{-5}\,\mathrm{rad}$, the distance to the crystal
    $a_{0} = 0.5\times 10^{-3}$ cm and the crystal length
    $L = 1\,\mathrm{cm}$. Right pane: The dependence of the emitted
    number of quanta on the beam angular spread
    $\Delta\theta_{\mathrm{e}}$ for the MAMI facility. The parameters
    of the electron beam are the same as on the left
    pane.}\label{fig:8}
\end{figure*}
In this section we present numerical results of the calculation of the
total number of photons $N_{\mathrm{PXR-SPG}}$ emitted by a single
electron in the Smith-Purcell geometry. We choose the most intense PXR
peaks for the $(400)$ and $(220)$ reflections in the Si crystal
\cite{SONES200522}. For these reflections the following parameters, taken from
the X-ray database \cite{StepanovXrayWebServer} are employed. For the
reflection $(400)$
\begin{equation}
  \begin{aligned}
    &\hbar\omega_{\mathrm{B}} = 6.45\,\mathrm{keV},& &k_{0} = 3.29\times10^{8}\,\mathrm{cm}^{-1},
    \\
    &\chi_{0}' = -0.24\times10^{-4},& &\chi_{0}'' = 0.83\times 10^{-6},
    \\
    &\chi_{g}' = 0.12\times 10^{-4},& &\chi_{g}'' = 0.71\times 10^{-6}
  \end{aligned}\label{eq:61}
\end{equation}
and $(220)$
\begin{equation}
  \begin{aligned}
    &\hbar\omega_{\mathrm{B}} = 4.51\,\mathrm{keV},& &k_{0} =
    2.3 \times 10^{8}\, \mathrm{cm}^{-1},
    \\
    &\chi_{0}' = -0.48 \times 10^{-4},& &\chi_{0}'' = 0.32 \times 10^{-5},
    \\
    &\chi_{g}' = 0.29 \times 10^{-4},& &\chi_{g}'' = 0.31\times 10^{-5}.
  \end{aligned}\label{eq:62}
\end{equation}
respectively. In addition we consider that the aperture of the
detector $\theta_{\mathrm{D}} = 10^{-2}$.

As a typical scenario we have investigated two accelerator facilities,
namely Mainz microtron MAMI \cite{Lauth2006, Brenzinger1997,
  PhysRevLett.79.2462}, where the typical electron beam energy
$\approx 900\,\mathrm{MeV}$ and SLAC Linac Coherent Light Source LCLS
\cite{LCLS_parameters,PhysRevLett.102.254801, PhysRevSTAB.11.030703},
where the electron beam energy varies from $2500\,\mathrm{MeV}$ to
$16900\,\mathrm{MeV}$. The MAMI facility provides the electron beam
with natural emittance
$\epsilon = 5\times 10^{-6}\,\mathrm{cm}\times \mathrm{rad}$, while
the normalized emittance on the LCLS facility
$\gamma\epsilon = (0.5-1.6)\times
10^{-4}\,\mathrm{cm}\times\mathrm{rad}$. For the MAMI accelerator we
assume that two different electron beams with similar emittances, but
different transverse sizes and angular spreads are employed in the
experiment, viz.
$\epsilon = 5\times 10^{-6}\,\mathrm{cm}\times \mathrm{rad}$ with
$\Delta \theta_{\mathrm{e}} = 10^{-2}$ rad,
$\Delta a = 5\times 10^{-4}$ cm and $\epsilon = 3\times 10^{-6}$
$\mathrm{cm}\times \mathrm{rad}$, with
$\Delta \theta_{\mathrm{e}} = 3\times10^{-3}$ rad,
$\Delta a = 10^{-3}$ cm.

In all following figures we plot the emitted number of quanta
normalized by the number of electrons, i.e., the total intensity is
divided by the number of electrons.

\subsection{The ideal case of a vanishing emittance}
\label{sec:ideal-case-vanishing}
We start a discussion from an ideal case when an electron beam does
not possess an emittance. In this situation the integration with a
distribution function
$G(\theta_{\mathrm{e}z},\theta_{\mathrm{e}y},z_{0})$ can be performed
analytically. For this we note that when $\Delta a \to 0$ and
$\Delta\theta_{\mathrm{e}} \to 0$ the distribution is localized near
point $z_{0} = a_{0}$. We now fix the value $a_{0} > 0$. In this
situation the only contribution to the integral is coming from the
region $[0,\infty)$. Consequently, we continue the integration to the
region $(-\infty,\infty)$ and integrate the differential number of
quanta (\ref{eq:57}) with the distribution
$\delta(z_{0} - a_{0}) \delta(\theta_{\mathrm{e}z})
\delta(\theta_{\mathrm{e}y})$. Performing this we are left with an
integral over $\rho d\rho d\varphi$. The integral in $\varphi$ is
trivially performed and we arrive to
\begin{align}
  N_{\mathrm{SP}}^{\mathrm{ideal}}
  &= \frac{k_{0} L}{2}
  \frac{e_{0}^{2}}{\hbar c}|\chi_g|^{2} \nonumber
  \\
  &\mspace{60mu}\times
    \int_{\gamma^{-1}}^{\sqrt{\theta_D^2+\gamma^{-2}}}dt
  \frac{t(t^2-\gamma^{-2})}{(t^2-\chi_0')^2}e^{-2k_0 a_0 t}.\label{eq:63}
\end{align}
This integral can be computed analytically and expressed through the
exponential integral function $\mathrm{Ei}(x)$. However, the
resulting expression is rather cumbersome and we do not list it
here.

The dependence on the main parameters can be deduced from
Eq.~(\ref{eq:63}). Thus, the emitted number of quanta is proportional
to the crystal length and exponentially decays with the increasing
value of the impact parameter $a_{0}$. For ultra-relativistic
electrons the dependence on the energy is weak and is given via
$\mathrm{const}+O(\gamma^{-2})$. Concluding, in Fig.~\ref{fig:9} we
plot the dependence of the total emitted number of quanta
$N_{\mathrm{SP}}^{\mathrm{ideal}}$ on the impact parameter $a_{0}$. As
can be seen from the figure $N_{\mathrm{SP}}^{\mathrm{ideal}}$ quickly
decays and for realistic impact parameters is negligible.

At the same time, when the impact parameter $a_{0} = 0$ the intensity
of SP can be comparable with the intensity of PXR-EAD
\cite{SKOROMNIK201786}, since both are $\sim 10^{-5}$ (see
Fig.~\ref{fig:9} and Ref.~\cite{SKOROMNIK201786}).

\subsection{The realistic situation of non-vanishing emittance}
\label{sec:real-situ-non}

We now introduce the restrictions coming from the realistic emittances
of experimental facilities. First of all, we compare the relative
contributions into the total radiation intensity
$N_{\mathrm{PXR-SPG}}$ from the parts of the electron beam that are
moving in crystal and in vacuum (pure SP radiation) respectively,
analogously to the ideal case. According to the qualitative estimation
(\ref{eq:14}) and Eq.~(\ref{eq:63}) of the ideal case the
Smith-Purcell radiation can be significant when the beam propagates
close to the crystal surface ($a_{0}$ is small) and possesses a small
transversal width $\Delta a$. Consequently, in Fig.~\ref{fig:4} we
plot the dependence of $N_{\mathrm{SP}}$ on $a_{0}$, when
$\Delta a = a_{0}/2$. As follows from the figure, for a realistic beam
size the relative contribution of this part is small, which agrees
with the qualitative estimation of Eq.~(\ref{eq:14}) and with the
ideal case investigated above.

Proceeding we would like to verify that since the value of
$N_{\mathrm{SP}}$ is small the main contribution to
$N_{\mathrm{PXR-SPG}}$ is coming from the tail of the electron beam
(beam halo), which is moving inside a crystal and, consequently, is
given via PXR-EAD. For this we plot in Fig.~\ref{fig:3} the results of
the numerical evaluation of the total number of photons
$N_{\mathrm{PXR-SPG}}$ emitted by a beam of electrons normalized by
the number of electrons as a function of the beam entrance angle
$\theta_{0}$. In addition we consider that the crystal length
$L\gg L_{\mathrm{abs}} = (k_{0}\chi_{0}'')^{-1}$, which is a realistic
scenario. In this situation we expect to observe a similar dependence
as in our previous work \cite{SKOROMNIK201786}. As follows from
Fig.~\ref{fig:3} the number of photons is significantly increasing
when the angle
$\theta_{0} < [Lk_{0}\chi_{0}'']^{-1} = L_{\mathrm{abs}}/L$, which
renders out hypothesis as the correct one.

As a result the dependence of $N_{\mathrm{PXR-SPG}}$ on various
parameters of the beam are analogous to the ones, which were
investigated in the Ref~\cite{SKOROMNIK201786}. In particular, in
Fig.~\ref{fig:5} we plot the dependence of $N_{\mathrm{PXR-SPG}}$ on
the beam energy, when the crystal length $L$ is fixed. The dependence
$N_{\mathrm{PXR-SPG}}(E)$ after reaching some $E_{\mathrm{opt}}$
saturates. The value $E_{\mathrm{opt}}$ is determined by the
parameters of the crystal.

In Fig.~\ref{fig:6} we plot the dependence of the number of emitted
quanta on the crystal length $L$, when the electron energy is fixed at
$E = 900$ MeV at MAMI and at $E = 8000$ MeV at LCLS. When the crystal
length increases to $L_{\mathrm{opt}}\approx 2$ cm the PXR intensity
saturates. For crystal lengths $L > L_{\mathrm{opt}}$ the intensity
slowly growths.

A specific peculiarity of PXR-SPG is a strong dependence of the
radiation intensity on the distance $a_{0}$ to the crystal
surface. Therefore, in Fig.~\ref{fig:7} we plot the
dependence on $a_{0}$. As follows from the figure, when the parameter
$a_{0}$ increases the intensity exponentially decreases.

With this we can conclude that practically under realistic
experimental conditions it is impossible to observe pure SP radiation,
since it is few order of magnitude less than the corresponding
PXR-EAD. Nevertheless, the geometry of PXR-SPG can be exploited for
the non-destructive diagnosis of the electron beam, due to its strong
dependence on an electron emittance that can be used as a complement
to the conventional knife-edge method
\cite{SMOLYAKOV200073,doi:10.1116/1.4802920}. For this reason, in
Fig.~\ref{fig:8} we show the dependence of the radiation intensity on
the parameters of the electron beam $\Delta a$ and
$\Delta\theta_{\mathrm{e}}$. It can be concluded that both components
of the emittance influence the radiation intensity, which can be
useful for the diagnosis of relativistic electron beams. It is
essential that the radiation is formed only by a small part of the
beam corresponding to the tail of its angular distribution and
consequently the characteristics of the majority of the electrons are
not changed.

\section{Conclusions}
\label{sec:conclusions}

In our work we have investigated the PXR radiation in the
Smith-Purcell geometry (PXR-SPG), when an electron beam propagates in
\textit{vacuum} parallel to the crystal-vacuum interface. We have
developed a theory, which describes this phenomenon and is based on
the dynamical diffraction theory. Our approach takes into account
peculiar features due to the grazing incidence of an electron beam on
a crystal surface. The reciprocity theorem allowed us to express the
spectral-angular distribution of the radiation intensity only through
the solutions of the homogeneous Maxwell equations.

We have demonstrated that the total number of photons
$N_{\mathrm{PXR-SPG}}$, emitted in the PXR peak is mainly given by the
electrons which are moving inside a crystal, while the pure SP
radiation is few order of magnitudes smaller than its PXR-EAD
counterpart.

At the same time, since the PXR-SPG radiation is mainly formed by the
beam halo, which moves inside a crystal and the total number of
emitted photons strongly depends on the parameters of an electron
beam, that is on both components of an electron emittance $\Delta a$
and $\Delta\theta_{\mathrm{e}}$, it can be used for
non-destructive diagnosis tool for relativistic electron beams.

\section{Acknowledgements}
\label{sec:acknowledgements}

ODS is grateful to C. H. Keitel for helpful discussions.

\appendix
\section{List of notations for electromagnetic waves}
\begin{itemize}
\item $\vec E_{\vec ks}^{(\pm)}(\vec r,\omega)$ --- Solutions of the
  homogeneous Maxwell equations for the ingoing $-$ (outgoing $+$)
  electromagnetic waves.
\item $E_{\vec ks}$ and $E_{\vec k_{g}s}$ --- Magnitudes of the direct
  and diffracted waves in a crystal respectively.
\item $\vec E^{(0)}_{\vec ks}$ --- Incident (direct) electromagnetic
  wave. $E^{(0)}_{\vec ks}$ its magnitude.
\item $\vec E_{\vec k_{g}s}^{(sp)}$ --- Specular reflected diffracted
  wave in vacuum. $E_{\vec k_{g}s}^{(sp)}$ its magnitude.
\item $E_{\mu s}$ and $E_{g\mu s}$ ($\mu = 1,2$) --- Magnitudes of the
  direct and diffracted waves inside a crystal, which correspond to
  the different solutions of the dispersion relation.
\end{itemize}

In addition, the diffraction of the wave with the vector $\vec k'$ in
Fig.~\ref{fig:1} happens on the crystallographic planes described by
the reciprocal vector $\vec g'$. At the same time, the diffraction of
the wave with the vector $\vec k = -\vec k'$ is happening on the
planes with the reciprocal vector $\vec g$ (see Fig.~\ref{fig:2}).




\bibliographystyle{apsrev4-1}
\bibliography{pxr_nim}



\end{document}